\documentclass[preprint2]{aastex63}

\defcitealias{Pineda2010}{P10}
\defcitealias{Pineda2011}{P11}
\defcitealias{Pineda2015}{P15}
\defcitealias{Schmiedeke2021}{S21}
\defcitealias{ChenM2020}{C20}

\usepackage{lipsum}  
\usepackage{amsmath}
\usepackage{gensymb}
\usepackage{textgreek}

\shorttitle{B5 Paper draft}
\shortauthors{Chen et al.}
\graphicspath{{./}{figures/}}

\begin{document}

\title{Turbulence and Accretion: a High-resolution Study of the B5 Filaments}

\correspondingauthor{Michael Chun-Yuan Chen}
\email{chen.m@queensu.ca} 

\author[0000-0003-4242-973X]{Michael Chun-Yuan Chen}
\affiliation{Department of Physics, Engineering Physics \& Astronomy, Queen's University, Kingston, Ontario, K7L 3N6, Canada}

\author[0000-0002-9289-2450]{James Di Francesco}
\affiliation{Herzberg Astronomy and Astrophysics, National Research Council of Canada, 5071 West Saanich Road, Victoria, BC, V9E 2E7, Canada}
\affiliation{Department of Physics and Astronomy, University of Victoria, Victoria, BC, V8P 1A1, Canada}

\author[0000-0002-3972-1978]{Jaime E. Pineda}
\affiliation{Max-Planck-Institut für extraterrestrische Physik, Giessenbachstrasse 1, 85748 Garching, Germany}

\author[0000-0003-1252-9916]{Stella S. Offner}
\affiliation{Department of Astronomy, University of Texas at Austin, Austin, TX 78712, USA}

\author[0000-0001-7594-8128]{Rachel K. Friesen}
\affiliation{Department of Astronomy \& Astrophysics at the University of Toronto, 50 St. George Street (Room 101) Toronto, Ontario, Canada, M5S 3H4}



\begin{abstract}
High-resolution observations of the Perseus B5 ``core'' have previously revealed that this subsonic region actually consists of several filaments that are likely in the process of forming a quadruple stellar system. Since subsonic filaments are thought to be produced at the $\sim 0.1$ pc sonic scale by turbulent compression, a detailed kinematic study is crucial to test such a scenario in the context of core and star formation. Here we present a detailed kinematic follow-up study of the B5 filaments at a 0.009 pc resolution using the VLA and GBT combined observations fitted with multi-component spectral models. Using precisely identified filament spines, we find a remarkable resemblance between the averaged width profiles of each filament and Plummer-like functions, with filaments possessing FWHM widths of $\sim 0.03$ pc. The velocity dispersion profiles of the filaments also show decreasing trends towards the filament spines. Moreover, the velocity gradient field in B5 appears to be locally well ordered ($\sim 0.04$ pc) but globally complex, with kinematic behaviors suggestive of inhomogeneous turbulent accretion onto filaments and longitudinal flows towards a local overdensity along one of the filaments.
\end{abstract}



\section{Introduction} \label{sec:intro}

Interstellar filaments appear to be intricately connected to star formation (\citealt{Schneider1979}; \citealt{Balsara2001}; \citealt{Andre2014}). Observationally, filaments are not only ubiquitous in molecular clouds (e.g., \citealt{Andre2010}; \citealt{Menshchikov2010}; \citealt{Miville-Deschenes2010}; \citealt{Hill2011}; \citealt{Roy2015}) they also seem to host the majority of the star-forming cores \citep{Konyves2015}. Indeed, filaments are not only naturally produced by supersonic turbulence in simulated molecular clouds (e.g., \citealt{Porter1994}; \citealt{Vazquez-Semadeni1994}; \citealt{Padoan2001}), the densest gas within these simulated filaments is also able to collapse and form stars under self-gravity (e.g., \citealt{Ostriker1999}; \citealt{Ballesteros-Paredes1999}; \citealt{Klessen2000}; \citealt{Bonnell2003}; \citealt{MacLow2004}; \citealt{Tilley2004}; \citealt{Krumholz2007}). Details on how filaments form and evolve in actual clouds, however, remain to be understood well, particularly in regards to how filaments assemble mass into star-forming cores.

Inferred from observations, filaments seem to possess a characteristic width of $\sim 0.1$ pc (e.g., \citealt{Arzoumanian2011}; \citealt{Palmeirim2013}; \citealt{Arzoumanian2019}). The origin of such a width has been attributed to the sonic scale at which point supersonic turbulent gas becomes subsonic, set by the cloud properties typically found in the Milky Way spiral arms \citep{Federrath2016}. Interestingly, this $\sim 0.1$ pc sonic scale also closely resembles that found empirically for dense cores \citep{Goodman1998}. Furthermore, subsonic levels of non-thermal motion have been observed in both dense cores (e.g., \citealt{Barranco1998}; \citealt{KirkHelen2007}; \citealt{Rosolowsky2008}) and filaments (e.g., \citealt{Pineda2011}; \citealt{Hacar2011}; \citealt{Hacar2016}), suggesting that cores and filaments likely share a common origin associated with such a sonic transition.

While the non-thermal motions exhibited by subsonic cores have often been attributed to turbulence (e.g., \citealt{Goodman1998}; \citealt{Pineda2010}; \citealt{ChenHope2019}), infall motions have been proposed lately as an alternative interpretation under the global hierarchical collapse scenario (hereafter, GHC; \citealt{Vazquez-Semadeni2019}). The turbulence interpretation, which often fits under the ``gravoturbulent'' scenario (hereafter, GT; e.g., \citealt{MacLow2004}), implies that subsonic cores are formed by shock compression followed by turbulence dissipation (e.g., \citealt{ChenHope2020}). The GHC infall interpretation, on the other hand, suggests cores originate as density enhancements in mildly supersonic clouds and only start to contract when the average Jeans mass in the cloud is reduced to that comparable to the cores as the result of ongoing cloud contraction \citep{Vazquez-Semadeni2019}. Both of these interpretations for dense core observations likely hold for subsonic filaments as well, particularly given that the two appear to be intimately linked.

The key to constraining core formation models well requires a detailed understanding of the non-thermal motions within cores and the kinematic relationship between cores, filaments, and their shared environments. For example, filaments formed out of magnetized shock-compressed layers are expected to accrete gas preferentially along magnetic field lines, which are not necessarily perpendicular to filaments and can be distorted by gravitationally-induced motions as filaments evolve (e.g., \citealt{ChenCheYu2014}). Filaments formed in the GHC scenario in the absence of a magnetic field, on the other hand, are expected to accrete material from their surroundings in a perpendicular direction and redirect these flows along their lengths in the parallel direction to feed star-forming cores and clumps (e.g., \citealt{Gomez2014}). Therefore, observing gas flows of actual filaments is invaluable to understanding the formation and evolution of cores and filaments.

Observational studies in the past have measured filament velocity gradients either in the perpendicular (e.g., \citealt{Palmeirim2013}; \citealt{Fernandez-Lopez2014}; \citealt{Dhabal2018}) or parallel (e.g., \citealt{KirkHelen2013}; \citealt{Friesen2013}) directions relative to a filament. The former and the latter measurements have often been interpreted as accelerating or decelerating accretion flows onto and along filaments, respectively. By using precisely identified filament spines as local references of orientations, \cite{ChenM2020} simultaneously measured velocity gradients in both the parallel and the perpendicular directions on the $\sim 0.05$ pc scale and revealed a wealth of complex velocity structures that is not necessarily expected from simple accretion scenarios.

In this paper, we expand on the methods developed by \citeauthor{ChenM2020} (\citeyear{ChenM2020}; hereafter \citetalias{ChenM2020}) to explore the gas kinematics of the Perseus B5 region at a $\sim 0.009$ pc resolution with the NH$_3$ (1,1) observations first obtained by \citeauthor{Pineda2015} (\citeyear{Pineda2015}). Since dense gas tracers such NH$_3$ and N$_2$H$^{+}$ can still trace multiple velocity components along lines of sight despite their low volume-filling fraction in the cloud (e.g., \citealt{Hacar2017}; \citetalias{ChenM2020}; \citealt{Choudhury2020}), we model each NH$_3$ (1,1) spectrum with up to two spectral components using the \texttt{MUFASA} software (\citetalias{ChenM2020}) to uncover a second NH$_3$ component previously unseen in the B5 region (\citealt{Pineda2010}; \citealt{Pineda2011}; \citealt{Pineda2015}; \citealt{Schmiedeke2021}; hereafter \citetalias{Pineda2010}, \citetalias{Pineda2011}, \citetalias{Pineda2015}, and \citetalias{Schmiedeke2021}). Even when a spectrum is dominated by a primary component, neglecting the presence of other detectable components can result in erroneous measurements of gas kinematics and structure dynamics (e.g., \citealt{Choudhury2020}).

At a distance of 302 pc \citep{Zucker2018}, Perseus B5 is a well-studied star-forming clump with the first known observation of a sharp spatial transition between a subsonic core and its supersonic surroundings (\citetalias{Pineda2010}). The clump is also known to host at least a single protostar (B5-IRS1; \citealt{Fuller1991}). A subsequent high-resolution ($\sim 0.009$ pc) study of B5 revealed that the B5 ``core'' actually consists of several filaments that are well embedded within the sonic region (\citetalias{Pineda2011}). A more sensitive followup of the \citetalias{Pineda2011} study at the same spatial resolution further revealed that regions of overdensity within these B5 filaments and the B5-IRS1 protostar seem to be gravitationally bound to each other and are likely in the process of forming a quadruple star system (\citetalias{Pineda2015}). These combined properties make B5 a highly attractive region to explore further the origin and evolution of cores and filaments on the verge of forming stars. 

Our paper is laid out as follows. In Section \ref{sec:data} and \ref{sec:method}, we briefly describe the B5 data and our analysis methods, respectively. We present our results in Section \ref{sec:results}, followed by a discussion of these results in Section \ref{sec:discussions}. We end our paper with a concluding summary in Section \ref{subsec:conclusion}.

\section{Data}\label{sec:data}

\begin{figure}
\centering
\includegraphics[width=0.92\columnwidth]{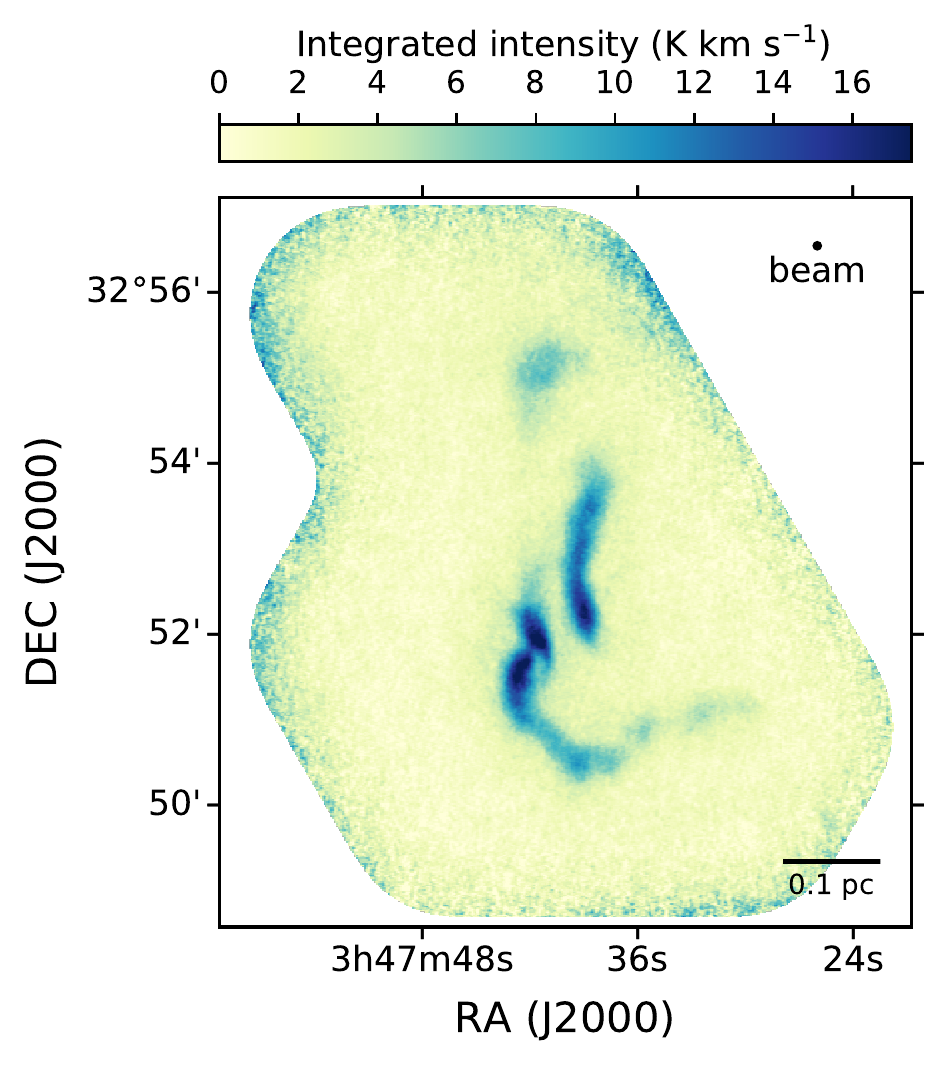}
\caption{The integrated intensity map of the NH$_3$ (1,1) emission seen towards the Perseus B5 region made with the VLA+GBT combined data. The beam corresponding to the data is also shown in the panel along with a physical scale bar. \label{fig:B5_mom0_map}}
\end{figure}

Our B5 data consist of interferometric observations of NH$_3$ (1,1) lines made by \citetalias{Pineda2015} with the Jansky Very Large Array (VLA) using the K-band receivers and the WIDAR correlator at a 3.90625 kHz channel width (i.e., 0.049 km s$^{-1}$ equivalent for the NH$_3$ spectra). The data were observed in the D-array and the CnD-array (project 11B-101) configurations over 27 mosaic pointings. To recover large-scale structures insensitive to the VLA, \citetalias{Pineda2015} further includes the single-dish NH$_3$ (1,1) observations of B5 taken by \citetalias{Pineda2010} with the Robert F. Byrd Green Bank Telescope (GBT). The GBT observations were made using the K-band and the On-The-Fly technique \citep{Mangum2007}. 

The combined VLA and GBT (VLA+GBT) data were reduced and imaged by \citetalias{Pineda2015} with the \texttt{CASA} software \citep{McMullin2007} using a 6$''$ circular beam with the multi-scale clean method and a robust parameter of 0.5, corrected by the primary beam. The final root mean square (rms) noise level of the reduced data is 0.24 K per channel, about 3.5 times lower than that of the \citetalias{Pineda2011} data, which is not a part of \citetalias{Pineda2015}'s reduced data product. For our analyses, we grid the final image over 1$''$ size pixels. At a distance of 302 pc, the physical sizes of the beam is $0.009$ pc. Figure \ref{fig:B5_mom0_map} shows the integrated intensity map of the final image from the VLA and GBT combined data.

\section{Method}\label{sec:method}

Our analyses build on robust measurements of kinematically continuous, i.e., coherent, structures. We accomplish such measurements by performing multi-component spectral fitting with statistical model selection and identifying filaments from the fits-derived emission models in the position-position-velocity (ppv) space. We describe these steps below in Section \ref{subsec:specfit} and Section \ref{subsec:fil_id}, respectively. Since the method required to reconstruct kinematically-coherent property maps based on multi-component fits is data-driven, i.e., depends on the type of information we can extract from the data, we will describe our reconstruction method later in the Results section (specifically, in Section \ref{subsec:vc_maps}).

\subsection{Spectral Fitting\label{subsec:specfit}}

We fit multi-component spectra to the NH$_3$ (1,1) data using the \texttt{MUFASA} software (\citetalias{ChenM2020}), which wraps around the \texttt{PySpecKit} package \citep{Ginsburg2011} to perform least-squares fitting via the Levenberg–Marquardt (LM; \citealt{Levenberg1944}; \citealt{marquardt1963}; \citealt{More1978}) method automatically. For each NH$_3$ (1,1) spectrum, we fit up to two velocity components per model, with each of these velocity components consisting of 18 hyperfine lines. We model each of our velocity components as a homogeneous gas slab parameterized by its velocity centroid ($v_{\mathrm{LSR}}$), velocity dispersion ($\sigma_v$), excitation temperature ($T_{ex}$), and optical depth ($\tau_0$). The final spectral model used in the fit is generated via radiative transfer that include the cosmic microwave background (CMB) through each individual slab along the line of sight towards the observer. We do not include the spectral response of the frequency channels in our model.

To ensure the automated fitting routine finds the global least-squares minima consistently and robustly, we adopt \texttt{MUFASA}'s default fitting recipe, which fits spectral cubes over two iterations (\citetalias{ChenM2020}). Under this recipe, \texttt{MUFASA} first spatially convolves the cube to twice its beamsize and then fits the convolved cube using initial parameter guesses derived from the cube's zeroth, first, and second moments. Following this initial fit, \texttt{MUFASA} then passes the fit results as initial guesses to a subsequent, full-resolution fit. This approach allows \texttt{MUFASA} to take advantage of the higher signal-to-noise ratio (SNR) of the convolved cube to assist with fits to the full-resolution cube. 

Once the full-resolution fits are obtained, \texttt{MUFASA} selects the best-fit model between the two-, one-, and zero-component (i.e., noise) fits on a pixel-by-pixel basis using the corrected Akaike Information Criterion (AICc; \citealt{Akaike1974}; \citealt{Sugiura1978}), where the small letter \textit{c} denotes the second-order correction made to account for finite sample sizes. \texttt{MUFASA} determines model \textit{b} to be the better model when the AICc-calculated relative likelihood of model \textit{b} over model \textit{a}, i.e.,
\begin{equation}\label{eq:lnk}
\ln{K_a^b} = - \left ( \textup{AICc}_b - \textup{AICc}_a \right )/2
\end{equation}
is above the statistically robust threshold of 5 \citep{Burnham2004}. This approach allows \texttt{MUFASA} to detect signals more robustly than adopting a simple SNR threshold and performs similarly to the Bayesian method implemented by \cite{Sokolov2020} while mitigating the need to sample likelihood spaces exhaustively for each fit. 

\texttt{MUFASA}'s ability to fit spectra and select the best-fit model correctly with its standard recipe is tested and presented by \citetalias{ChenM2020} based on fits to 30,000 randomly generated synthetic spectra. To improve our fits further beyond those provided by the standard recipe, we perform an additional iteration of fitting to the original data with initial guesses derived from the interpolated results obtained originally from the standard fits. Pixels with Jacobian-estimated $v_{\mathrm{LSR}}$ and $\sigma_{v}$ errors larger than $0.2$ km s$^{-1}$ are removed from the initial guesses before they are interpolated and smoothed with a Gaussian kernel with a width of $\sigma_{\mathrm{ker}} = 1$ pixel. For pixels with $\ln{K^1_0} < 50$, where the SNRs of spectra are expected to be lower, a kernel with $\sigma_{\mathrm{ker}} = 3$ pixels is used instead. The new iteration of fitted results is compared with the initial iteration on a per-pixel basis and only the best-fitted model as determined by the AICc is adopted. This additional iteration successfully improves the situation for pixels that are initially fitted poorly and scattered sparsely throughout regions that appear to be well-fitted otherwise, as examined by eye. We note this additional iteration merely provides more robust initial guesses for the least-squares fitting process and does not superficially dictate the final, best-fit results.

\subsection{Filament Identification\label{subsec:fil_id}}

Following \citetalias{ChenM2020}, we use the \texttt{CRISPy} software to identify filament spines in position-position-velocity (ppv) space from the best-fit spectral models with the hyperfine structures removed in post-processing. We remove hyperfine structures, which are not kinematic in nature, to reveal the true kinematic structures behind the observed spectra. Such a removal is accomplished by reconstructing each best-fit velocity component's optical depth (i.e., $\tau$) profile as a single Gaussian. To mimic the optical depths of the satellite hyperfine lines that are typically thin, we set the peak $\tau$ value of each velocity component to one-tenth the peak $\tau$ value of all the 18 hyperfine lines combined (i.e., $\tau_0$), as derived from the fits. To ensure the reconstructed structures are spectrally well-resolved for filament identification, we further adopted a constant, minimally Nyquist-sampled value of $\sigma_v = 0.043$ km s$^{-1}$ in our reconstruction instead of the values we derived from our best fits. We call such a reconstruction deblending and we only use the deblended emission cube for the purpose of filament identification. The $\sigma_v$ values derived from the best fits are still used in our final analyses.

We identified filament skeletons from the deblended cube using the \texttt{CRISPy} software developed by \citetalias{ChenM2020}, which finds these in multi-dimensional images as one-dimensional density ridges using the generalized Subspace Constrained Mean Shift (SCMS) method (\citealt{ChenYC2014arXiv}). The \texttt{Python}-based \texttt{CRISPy} software is derived and generalized from the \texttt{R} code written earlier by \cite{ChenYC2015MNRAS}. Filament skeletons identified through the SCMS framework have mathematically well-defined orientations set by the local density field. We further prune skeletons into branchless spines using the infrastructure developed by \citetalias{ChenM2020} based on its two-dimensional predecessor by \cite{Koch2015}. The filament spines are defined by the longest path within their parental skeleton. We note that filament spines identified with \texttt{CRISPy} in ppv space are kinematically continuous by definition.

\section{Results} \label{sec:results}

\subsection{Fitted spectra}

\begin{figure*}
\centering
\hspace{1.5mm}
\includegraphics[width=0.48\textwidth]{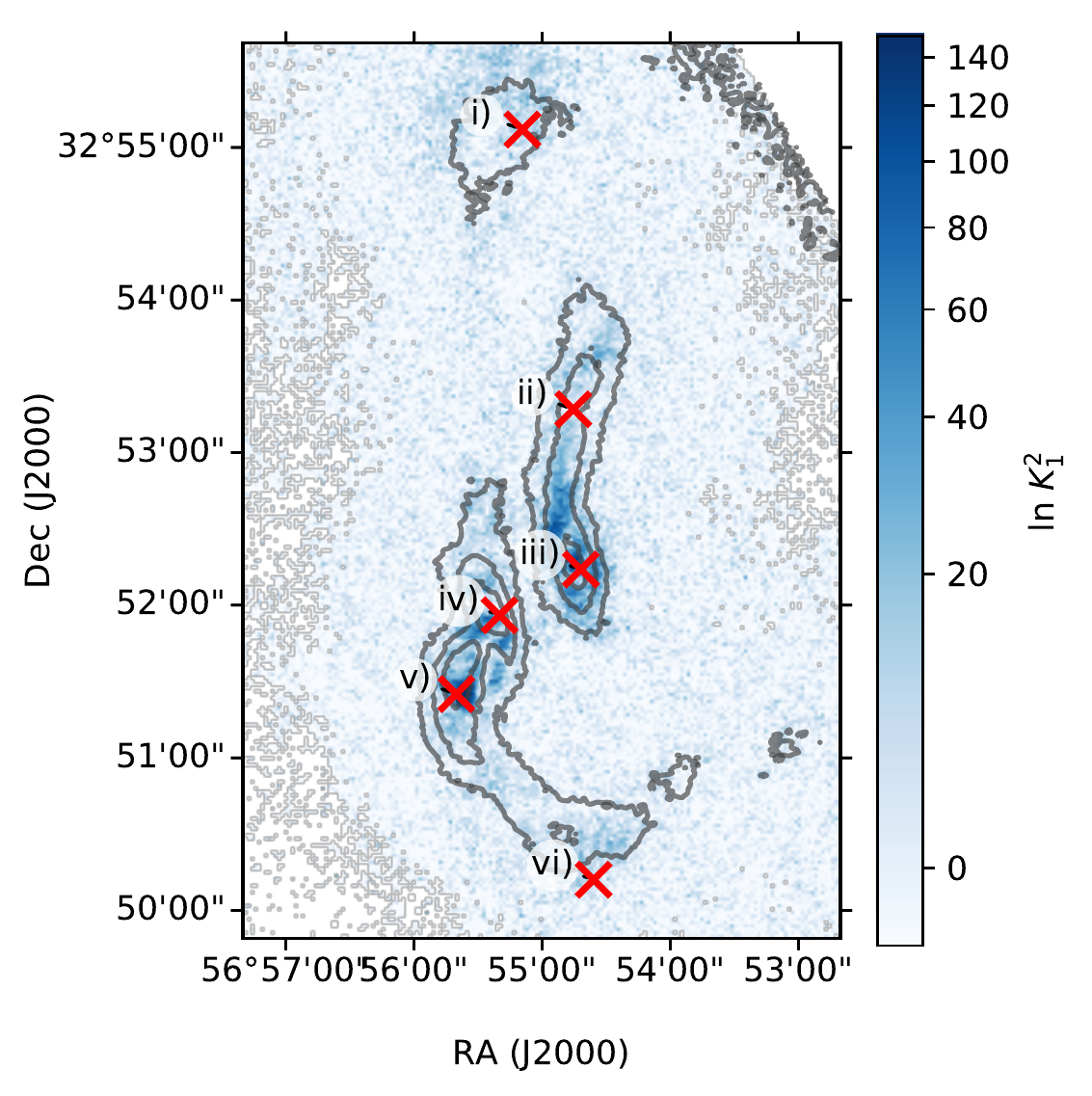}
\includegraphics[width=0.424\textwidth]{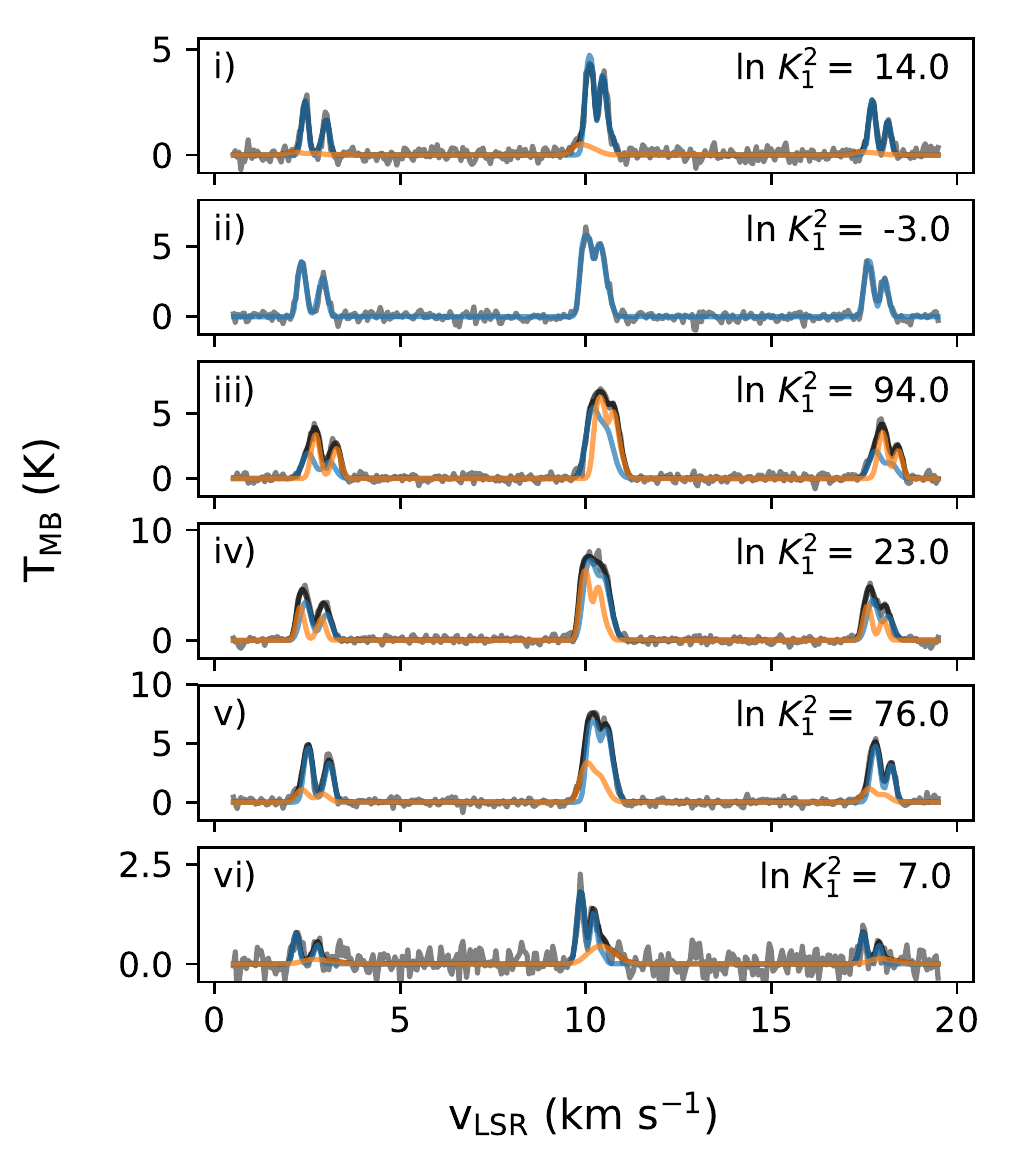}
\caption{The left panel shows the relative likelihood maps of our two-component fits over the one-component fits (i.e., $\ln{K^2_1}$) as determined from the AICc. Grey contours correspond to integrated intensities of 5 K km s$^{-1}$, 10 K km s$^{-1}$ and 15 K km s$^{-1}$. Red `x' symbols mark from where the example fits (black), their individual components (blue and orange), and their corresponding spectra (grey) shown in the right panels are taken.  \label{fig:samp_spec_VLA}}
\end{figure*}

The left panel of Figure \ref{fig:samp_spec_VLA} shows the relative log-likelihood map of the two-component fits to the B5 data over the one-component fits, as determined by the AICc, while the right panel shows example spectra and their respective best-fit models taken from positions marked in the left panels. Out of the 166,920 pixels within the footprint of the VLA+GBT data, we detected 119,398 spectra (i.e., $\ln{K^1_0} > 5$). Of these detected spectra, 19,800 pixels (i.e., 17.0\%) are better fitted with two-component models (i.e., $\ln{K^2_1} > 5$).

\subsection{Identified Filaments \label{subsec:id_fils}}

\begin{figure*}[t!]
\centering
\includegraphics[width=0.93\textwidth]{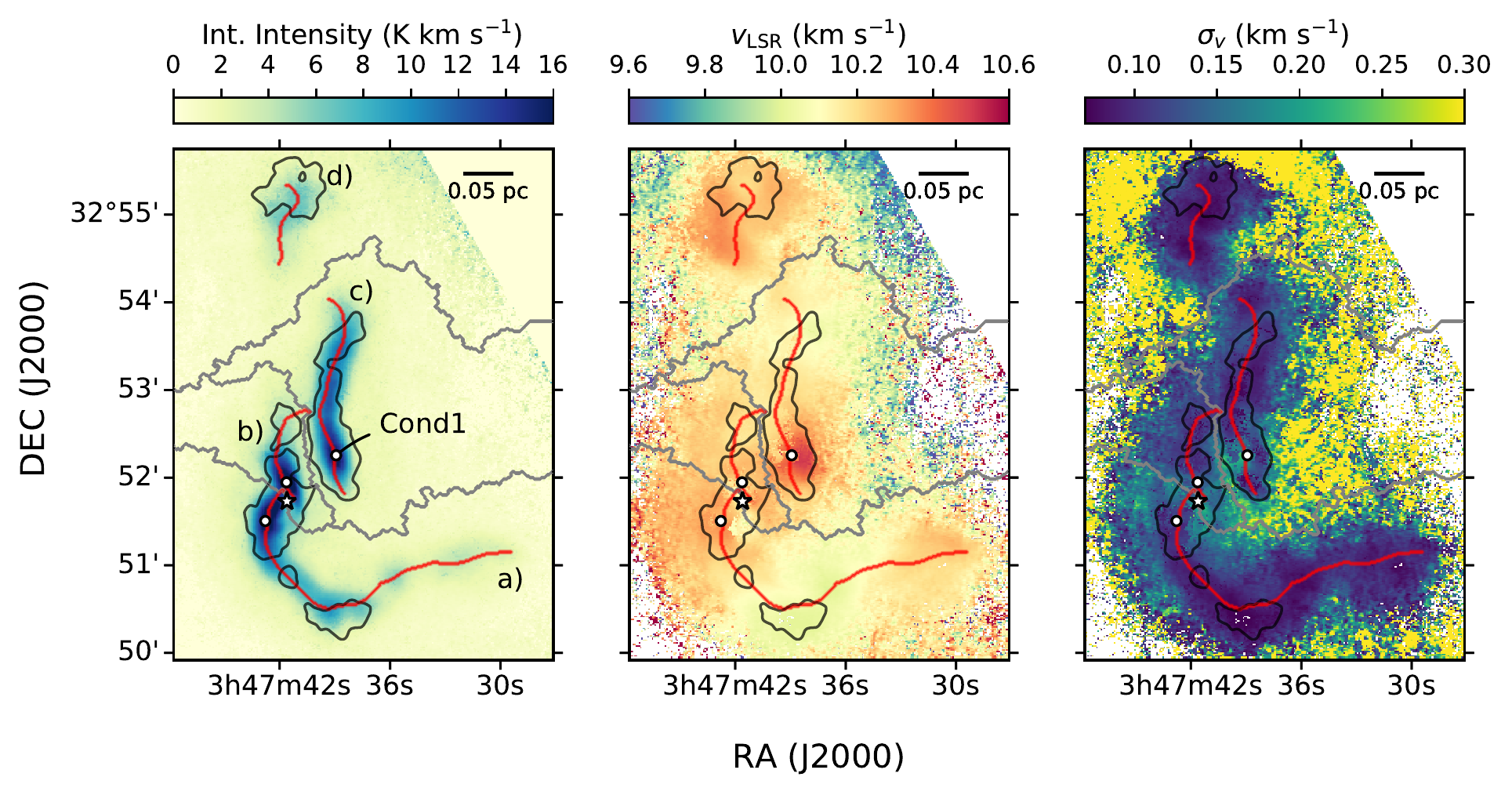}
\caption{Filament spines (red) identified by \texttt{CRISPy} overlaid on the total modelled integrated intensity (left), the sorted $v_{\mathrm{LSR}}$ (center), and the sorted $\sigma_v$ (right) maps of our best fit model to the NH$_3$ VLA data. The position of the IRS1 protostar and condensations identified by \citetalias{Pineda2015} are marked by the star and circles, respectively. The borders of the watershed regions, identified from the integrated intensity map, are shown in light grey. Smoothed contours of where the data are better fitted by two components, i.e., where $\ln{K^2_1} > 5$, are shown in black. Physical scale bars are shown in all panels. \label{fig:spine_maps}}
\end{figure*}

The left panel of Figure \ref{fig:spine_maps} shows the \texttt{CRISPy} identified filament spines overlaid on top of the integrated intensity map of our modelled fits to the VLA+GBT data. The branches initially identified by \texttt{CRISPy} as a part of filament skeletons have been removed, leaving the longest paths through these skeletons as our final, branchless spines. In B5, we identify four filament spines in total. They are named alphabetically from \textit{a} to \textit{d}, ordered from the filament furthest from the origin of the image, i.e., the southeastern corner, to the one that is the nearest. While filament \textit{d} appears by eye to have a relatively low aspect ratio, it does contain a well-defined density ridge traced by \texttt{CRISPy}. To ensure we have a complete, relatively unbiased sample, we include filament \textit{d} in our analyses.

The filaments we identify in B5 with \texttt{CRISPy} differ a bit in projection from those identified by \citetalias{Schmiedeke2021} in 2D from the integrated intensity map of the same data with \texttt{FilFinder} \citep{Koch2015}. While both methods identified filament \textit{c} (i.e., \textit{B5-Fil1}) as the same structure, \texttt{FilFinder} identified filament \textit{b} and the eastern half of filament \textit{a} as a single structure (i.e., \textit{B5-Fil2}) instead of two. The differences between these results suggest that \texttt{CRISPy} is better than \texttt{FilFinder} at detecting filaments over a higher dynamic range of brightness. In particular, the \texttt{FilFinder} results obtained by \citetalias{Schmiedeke2021} did not recognize the valley between filament \textit{a} and \textit{b}, which is brighter than the average background in B5. 

Since the four filamentary structures identified by \texttt{CRISPy} in PPV space do not overlap on the plane of the sky, we further divided B5 into topologically distinct regions based on the integrated intensity map of our best-fit NH$_3$ model using the watershed method implemented in the \texttt{scikit-image} package \citep{vanderWalt2014}. The watershed method segments multi-dimensional structures with isosurfaces that are shared by density peaks or `seeds,' and have been widely adopted by algorithms such as the \texttt{CLUMPFIND} \citep{WilliamsJ1994} and \texttt{CPROPS} \citep{Rosolowsky2006} to identify molecular cloud structures. To ensure the B5 clump is divided into regions defined by their respective filaments, rather than just any emission peak, we seeded the watershed segmentation with our identified filament spines. The boundaries resulting from the segmentation are indicated in Figure \ref{fig:spine_maps} in grey. 

\subsection{Kinematically-coherent Maps \label{subsec:vc_maps}}

To show better where two-component spectra are well detected, we plot black contours in Figure \ref{fig:spine_maps} of where the smoothed $\ln{K^2_1}$ map has a value of 5. We note the original, non-smoothed $\ln{K^2_1} = 5$ contours are rather noisy in fainter regions, which suggests the second component detected sparsely in these regions is likely ubiquitous over the observed footprint of B5 and is only detected marginally due to low SNRs. The one-component emission detected outside of the $\ln{K^2_1} = 5$ contours, however, appears to have fits-derived $v_\mathrm{LSR}$ and $\sigma_v$ maps that are rather smooth across the entire B5 region, which suggest the B5 regions contains a single, dominant structure that is kinematically coherent.

Since the one-component-fit-derived $v_\mathrm{LSR}$ and $\sigma_v$ maps appear to trace the dominant emission that is also detected in regions with $\ln{K^2_1} > 5$ (i.e., within the contours) as one of the two fitted components, we opt for using the one-component-derived $v_\mathrm{LSR}$ and $\sigma_v$ maps as the base reference to determine which one of the two components (when justified) belongs to the kinematically-coherent structure traced elsewhere by one component fits. We adopt the component that is kinematically the most similar to its neighboring pixels with one-component fits into our final maps for kinematic analyses. We note such an adoption merely isolates one of the two detected components for analyses and does not alter the results drawn from the already-fitted two-component models. 

We accomplished such an adaptation first by interpolating the $v_{\mathrm{LSR}}$ and $\sigma_v$ maps derived from one-component fits over pixels of two-component fits as if they were empty. We performed such an interpolation with a Gaussian kernel that is $\sigma_{\mathrm{ker}} = 0.25$ pixels in size to avoid degrading the image resolution significantly. Once we obtained the interpolated one-component map as the kinematically-coherent reference, we determine the kinematic similarities between the reference maps and each of the components by first calculating the differences between these for the $v_{\mathrm{LSR}}$ and $\sigma_v$ maps. We then use the quadrature sum of these difference maps as our metric of kinematic similarity.

The component determined to be the most similar to the reference maps is subsequently integrated, i.e., adopted, into the one-component maps to form a final set of $v_{\mathrm{LSR}}$ and $\sigma_v$ maps. The center and right panels of Figure \ref{fig:spine_maps} show these final $v_{\mathrm{LSR}}$ and $\sigma_v$ maps, respectively, overlaid with the same contours and symbols found in the left panel. The component we excluded from these final maps typically has estimated $v_{\mathrm{LSR}}$ errors that are at least twice those of their included counterparts, and has optical depths that are at least twice lower. Moreover, the $v_{\mathrm{LSR}}$ of the excluded component tends to form compact structures that are discontinuous on the $10''$ scale, making them unsuitable for velocity gradient analysis. We henceforth exclude these kinematically less similar components from our analyses. We note both the included and excluded components are derived from two-component fits still for pixels with $\ln{K^2_1} > 5$. We present further details of the excluded component in Appendix \ref{appendix:vcomps}.

\begin{figure}[t!]
\includegraphics[width=1.0\columnwidth]{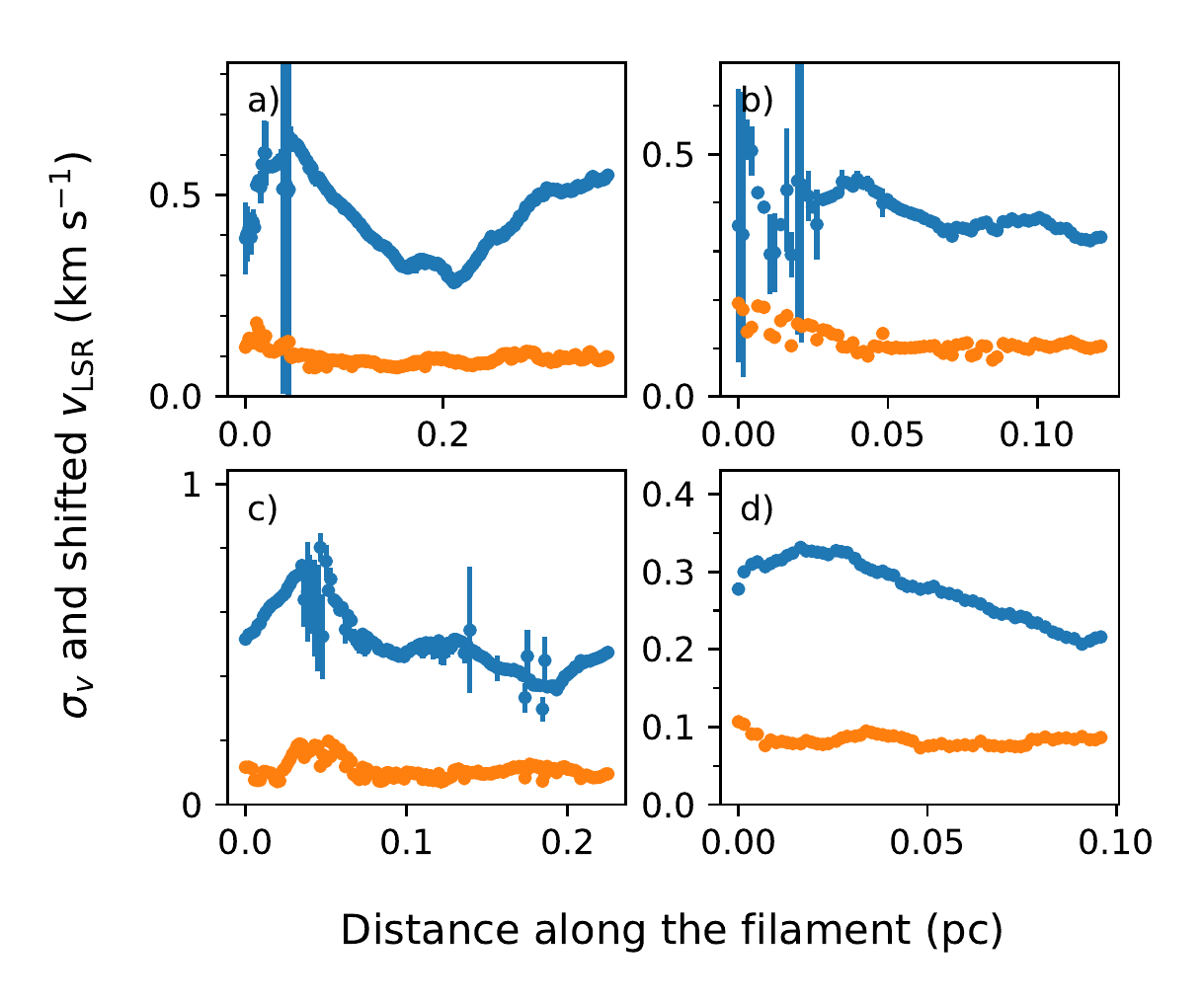}
\caption{The shifted $v_{\mathrm{LSR}}$ (blue) and $\sigma_v$ (orange) values of our final parameter maps along the four filament spines. The zero-point distance reference of the spine starts at the end closest to southeastern corner of the map, while the zero-point of the $v_{\mathrm{LSR}}$ shown is shifted arbitrarily to fit well within each panel. \label{fig:spine_profiles}}
\end{figure}

Figure \ref{fig:spine_profiles} shows the (shifted) $v_{\mathrm{LSR}}$ and the $\sigma_v$ taken from our final maps along the \texttt{CRISPy}-identified spines. With errors taken into consideration, the final $v_{\mathrm{LSR}}$ and $\sigma_v$ values are reasonably continuous along the spines, indicating that we are indeed tracing kinematically-coherent structures along their lengths. The abrupt increase in the estimated error in some instances results from both components in a two-component spectrum having similar $v_{\mathrm{LSR}}$ and $\sigma_{v}$ values, with at least one of them being optically thick. Under these circumstances, the optically thicker component closer to the observer can easily eclipse the rear component, obscuring the kinematic information encoded in the shape of the spectrum. Indeed, inspections of the fitted spectra confirm that fits with exceptionally large errors tend to have larger optical depths and similar $v_{\mathrm{LSR}}$ or $\sigma_{v}$ values between the two components. Moreover, pixels with large errors are preferentially found towards the centers of the densest structures (e.g., ridges), where the individual $\tau_0$ values of both components often exceed a value of 5. These indicators further suggest that high optical depths are the primary driver behind the large errors. 

\section{Discussion} \label{sec:discussions}

In this section, we present and discuss our analyses of the kinematically-coherent structure in our observations of B5, starting with an overview of its subsonic filaments in Section \ref{subsec:denseStructures}. We further discuss the structural and kinematic radial profiles of the filaments in Section \ref{subsec:rad_pro}, accompanied by a detailed look of B5's highly complex velocity gradients in Section \ref{subsec:vgrads}.

\subsection{Subsonic Filaments \label{subsec:denseStructures}}

As revealed by \citetalias{Pineda2011} and  \citetalias{Pineda2015} earlier, the dense, subsonic region in B5 is indeed not a monolith. In fact, the subsonic region commonly referred to as a `core' actually consists of several filaments. With \texttt{CRISPy} (\citealt{ChenM2020}), we identified four distinct filaments in B5 in the position-position-velocity space and designated them as \textit{a} to \textit{d} (see Figure \ref{fig:spine_maps}). 

While the B5 filaments have been referred to as `fibers' (e.g., \citealt{Andre2014}), a term first introduced by \cite{Hacar2013} to describe kinematically distinct but spatially unresolved bundles of sub-filaments, we will refer to the B5 substructures simply as filaments in this work. We note the fibers observed by \cite{Hacar2013} in Taurus with C$^{18}$O may not necessarily correspond to physical, three-dimensional filaments due to the complex gas kinematics often traced with species such as CO (e.g., \citealt{Zamora-Aviles2017}; \citealt{ClarkeS2018}). Moreover, although the partial, spatial overlap between filaments \textit{b} and \textit{c} along the east-west direction may resemble those identified spectrally by \cite{Hacar2013} in C$^{18}$O observations, such a spectral overlap is not typically found with denser gas tracers, such as NH$_3$ and N$_2$H$^+$ (e.g., \citealt{Tafalla2015}, \citealt{Hacar2017}, \citealt{Hacar2018}, \citealt{ChenM2020}).

About half (i.e., $51\%$) of the pixels from our final $\sigma_v$ map have values $<0.2$ km s$^{-1}$, indicating the gas at these pixels has subsonic non-thermal motions if they have a gas temperature of 10 K, a typical value measured by \citetalias{Schmiedeke2021} with NH$_3$ in B5 with one-component fits. This sonic estimate follows from when the non-thermal component of $\sigma_v$, i.e.,
\begin{equation}\label{eq:sigma_NT}
\sigma_{v, \mathrm{NT}} = \sqrt{\sigma_v^2 - k_b T_{\mathrm{kin}}/\mu_{_\mathrm{NH_3} m_{_\mathrm{H}}}}
\end{equation}
is less than the isothermal sound speed of a gas,
\begin{equation}\label{eq:SoundSpeed}
c_s = \sqrt{k_b T_{\mathrm{kin}}/\mu_{_\mathrm{ISM} m_{_\mathrm{H}}}},
\end{equation}
the total $\sigma_v$ have values of
\begin{equation}\label{eq:sonic_bound}
\sigma_v < \left[ \frac{k_b T_{\mathrm{kin}}}{m_{_\mathrm{H}}} \left ( \frac{1}{\mu_{_\mathrm{ISM}}} + \frac{1}{\mu_{_\mathrm{NH_3}}} \right ) \right]^{1/2}.
\end{equation}
Here, $k_b$ and $m_H$ are the Boltzmann constant and atomic hydrogen mass, respectively. If we assume a kinematic gas temperature ($T_{\mathrm{kin}}$), an NH$_3$ molecular weight ($\mu_{_\mathrm{NH_3}}$), and a mean interstellar molecular gas weight ($\mu_{_\mathrm{ISM}}$) of 10 K, 17.031, and 2.33, respectively, then the total $\sigma_v$ of a gas with a subsonic level of non-thermal motion should have values of $\lesssim 0.2$ km s$^{-1}$.
We note the thermal motion of a gas under such assumptions has a speed of $0.19$ km s$^{-1}$, which is marginally smaller than the total $\sigma_v$ due to $\mu_{_\mathrm{NH_3}} \gg \mu_{_\mathrm{ISM}} $.

For pixels within a $35''$ (i.e., 0.05 pc) radius of the filament spines, i.e., the typical filament half-widths commonly reported by Herschel studies (e.g., \citealt{Arzoumanian2011}), the fraction of pixels with $\sigma_v < 0.2$ km s$^{-1}$ is $83\%$. This result is consistent with the findings of \citetalias{Pineda2011} and \citetalias{Schmiedeke2021} in B5, where they measured $\sigma_v$ with one-component NH$_3$ fits over a smaller area.

\subsection{Radial Profiles \label{subsec:rad_pro}}

\begin{figure*}
\centering
\includegraphics[width=0.83\textwidth]{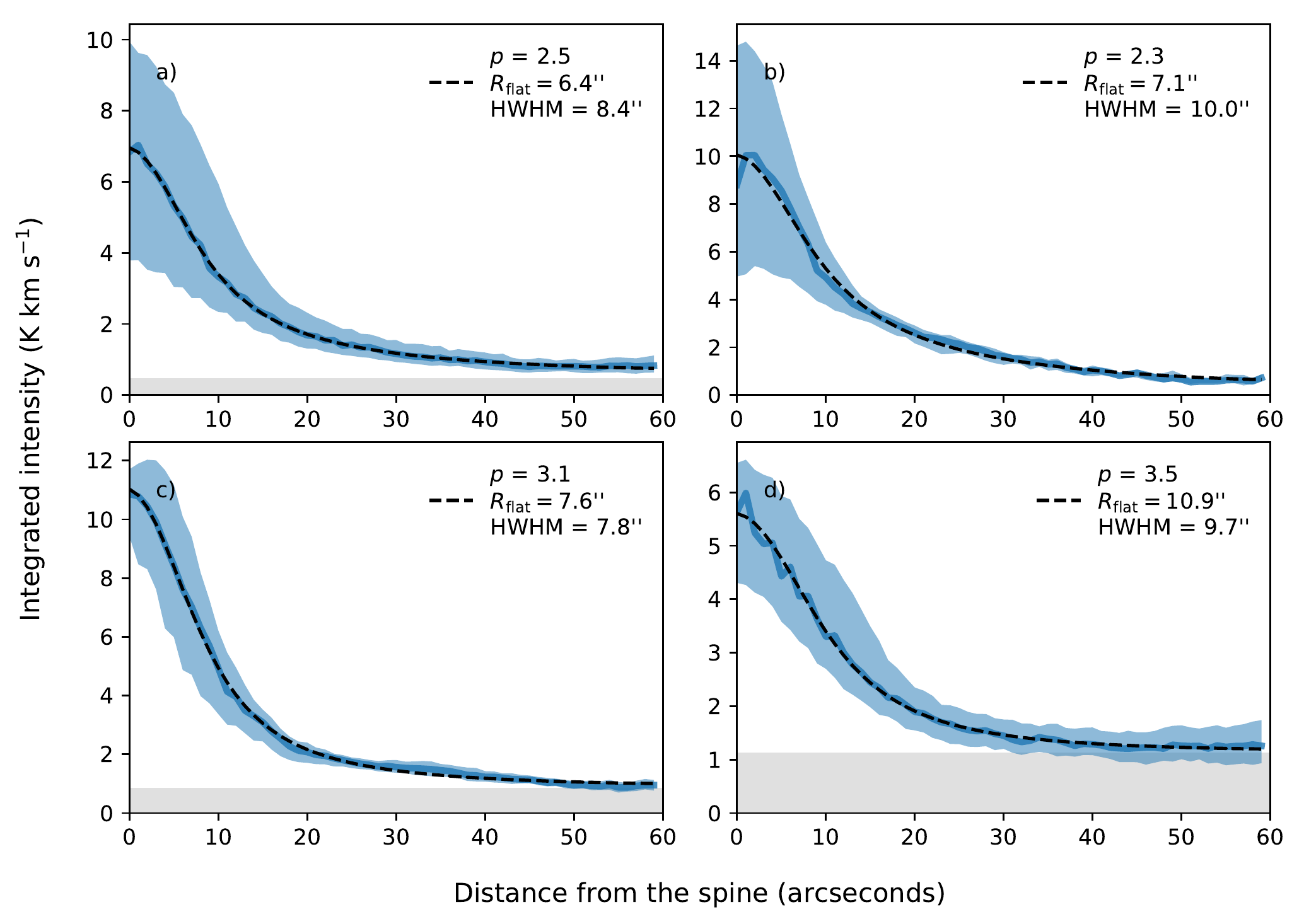} 
\caption{Averaged radial profiles of the NH$_3$ integrated intensity for the B5 filaments. The blue solid lines show the median value of the profiles while the blue shaded regions represent the 25- and 75-percentile ranges of the profile. The black dashed lines represents the best-fit Plummer-like models. \label{fig:radPro}}
\end{figure*}

\subsubsection{Radial integrated intensity profiles \label{subsub:radial_mom0}}

Figure \ref{fig:radPro} shows the averaged NH$_3$ integrated intensity radial profiles of all four B5 filaments measured along the direction perpendicular to the filament spines. The solid lines show the median value of the profile while the shaded region represents the 25- and 75-percentile ranges. We include only pixels within the watershed boundaries of each filament for their respective profiles to ensure contributions from neighbouring, unrelated filaments along each line of sight are minimized (see Section \ref{subsec:id_fils}). Such a practice is important for filaments that are close in proximity to each other on the plane of the sky, particularly given that filaments tend to have non-negligible, power-law-like profiles at larger radii (e.g., \citealt{Arzoumanian2011}).

We fitted radial integrated intensity profiles of each filament with a Plummer-like functions (see \citealt{Plummer1911}; \citealt{Whitworth2001}; \citealt{Nutter2008}; \citealt{Arzoumanian2011}), i.e.,

\begin{equation}\label{eq:Plummer}
I\left ( r \right ) = I_0 \left [ 1 + \left (r/ R_{\mathrm{flat}} \right )^2 \right ]^{\frac{1-p}{2}} + I_{\mathrm{bg}},
\end{equation}
where $r$ is the radial distance from the filament spine, $R_{\mathrm{flat}}$ is the radius of the flat inner region, and $p$ is the power-law index of the outer region. The $I_0$ and $I_{\mathrm{bg}}$ are the integrated intensities of the filament spine and the background emission, respectively. To emulate our observations, we convolved the Plummer-like functions with a $6''$ full width at half maximum (FWHM) Gaussian beam prior to fitting. To interpret integrated intensity as a proxy for gas column density, we assume the observed NH$_3$ emission is optically thin and does not have widely varying temperature. In the case where the best-fit $I_{\mathrm{bg}}$ value is zero, i.e., the lower limit constrain of our model, we refit the Plummer-like profile with $I_{\mathrm{bg}}$ fixed at zero as a constant to determine the covariance matrix of our fit for error estimates. 

The Plummer-like functions that best fit integrated intensity radials profiles of filaments in B5 are overlaid in Figure \ref{fig:radPro}. The best-fit $p$ values for filaments \textit{a}, \textit{b}, \textit{c}, and \textit{d} are $2.5 \pm 0.1$, $2.3 \pm 0.1$, $3.1 \pm 0.1$, and $3.5 \pm 0.4$, respectively. 

\citetalias{Schmiedeke2021} also made similar radial profile measurements using the same VLA+GBT observations of B5. Specifically, they made measurements for filaments they identified as \textit{B5-fil1} and \textit{B5-fil2}, using only the integrated intensity map in the position-position space. By fitting Plummer-like functions to profiles averaged across the entire filaments, \citetalias{Schmiedeke2021} measured $p$ values of $2.91 \pm 0.06$ and $2.98 \pm 0.05$ for \textit{B5-fil1} and \textit{B5-fil2}, respectively. The $p$ value they measured for \textit{B5-fil1}, i.e., our filament \textit{c}, is very consistent with the value we measured. The $p$ value they measured for \textit{B5-fil2}, on the other hand, is larger than those we find in filaments \textit{a} and \textit{b}, its rough counterparts. We note that \citetalias{Schmiedeke2021} identified filaments \textit{b} and half of filament \textit{c} as one single filament, which is likely the primary driver behind the apparent discrepancy between our measurements and theirs. 

The $p\geq 3$ Plummer-like fits we find for filaments \textit{c} and \textit{d} are higher than those generally found in studies conducted with dust continuum (e.g., \citealt{Arzoumanian2011}) and CO emission (e.g., \citealt{Panopoulou2014}; \citealt{Suri2019}) observations, which found profiles that are better fitted with $p=2$ than $p=4$ functions. Similarly, studies that fit Plummer-like function with a free-floating $p$ also find best-fit $p$ values to be closer to 2 rather than 4, both in observations (e.g., \citealt{Arzoumanian2019}) and simulations (e.g., \citealt{Smith2014}). While dense gas tracers, as pointed out by \cite{Arzoumanian2019}, can in principle have steeper power-law wings that give rise to higher $p$ values due to their insensitivity to more diffuse, lower-density gas and dust, such an interpretation does not fully explain why filament \textit{a} and \textit{b} do have best-fit $p$ values closer to 2.

The FWHM filament widths we measure from our best-fit radial profiles of filaments \textit{a}, \textit{b}, \textit{c}, and \textit{d} are $0.025 \pm 0.01$ pc, $0.029 \pm 0.02$ pc, $0.023 \pm 0.01$ pc, and $0.028 \pm 0.03$ pc, respectively. While these widths are significantly smaller than the $\sim 0.1$ pc widths commonly found with dust emission studies (e.g., \citealt{Arzoumanian2011}; \citealt{Andre2016}; \citealt{Arzoumanian2019}), they are nevertheless very similar to those found with ALMA observations of dust continuum emission in IRDC G035.39-00.33 ($\sim 0.028$ pc; \citealt{Henshaw2017}), as well as VLA NH$_3$ ($\sim 0.02$ pc; \citealt{Monsch2018}) and ALMA N$_2$H$^+$ ($\sim 0.035$ pc; \citealt{Hacar2018}) observations of Orion A. We note that unlike many of the dust continuum studies, we measure our FWHM filament widths directly from the fitted function, i.e., nonparametrically, rather than inferring them from the fitted $R_{\mathrm{flat}}$ parameter through a power-law-dependent scaling factor. 

Similar to the ALMA dust continuum and N$_2$H$^+$ emission measurements made by \cite{Henshaw2017} and \citealt{Hacar2018}, respectively, the $\sim 0.03$ pc FWHM filament widths we find in B5 are about a factor of three smaller than the $\sim 0.1$ pc width typically found with single-dish, dust continuum studies, which have spatial resolutions $\sim 3-4$ times lower than that of the VLA data (e.g., \citealt{Arzoumanian2011}; \citealt{Andre2016}; \citealt{Arzoumanian2019}). The fact that the $\sim 0.03$ pc widths have been observed in dust continuum studies, with extended emission recovered (\citealt{Henshaw2017}), suggests that the $\sim 0.03$ pc widths measured with denser gas tracers such as NH$_3$ and N$_2$H$^+$ may not be necessarily biased significantly by the smaller volumes these molecular species trace relative to those traced by dust.

Indeed, the HWHM widths measured by \cite{Priestley2020} from synthetic observations of their fiducial simulations with post-processed chemical networks have comparable values between those measured with NH$_3$, 850 \textmu m dust emission, and the actual gas column density. While their median HWHM width measured nonparametrically with NH$_3$ is 0.12 pc, much higher than those we have measured, \cite{Priestley2020} acknowledged that their samples often contain double peaks and their profiles have narrower widths if measured from careful fitting of Plummer-like profiles. Interestingly, the HWHM widths they measured with N$_2$H$+$ are about half of those measured with NH$_3$ in the same model.

Even without synthetic observations, many simulations have also shown filament widths in the $0.02-0.04$ pc range (e.g., \citealt{Juvela2012}; \citealt{ChenCheYu2020}; \citealt{Heigl2020}). Given the filaments we identified in B5 are all less than 0.4 pc in length, with two having spine lengths of $\sim 0.1$ pc, the long aspect ratio definition of filaments necessitates that these structures must have widths significantly narrower than the commonly reported 0.1 pc in observations. Such constraints further support that the narrow filament widths we observed in B5 are indeed real, even if they are somewhat biased by denser gas tracers.

\subsubsection{Radial kinematic profiles. \label{subsub:radial_vel}}

\begin{figure}
\centering
\includegraphics[width=0.45\textwidth]{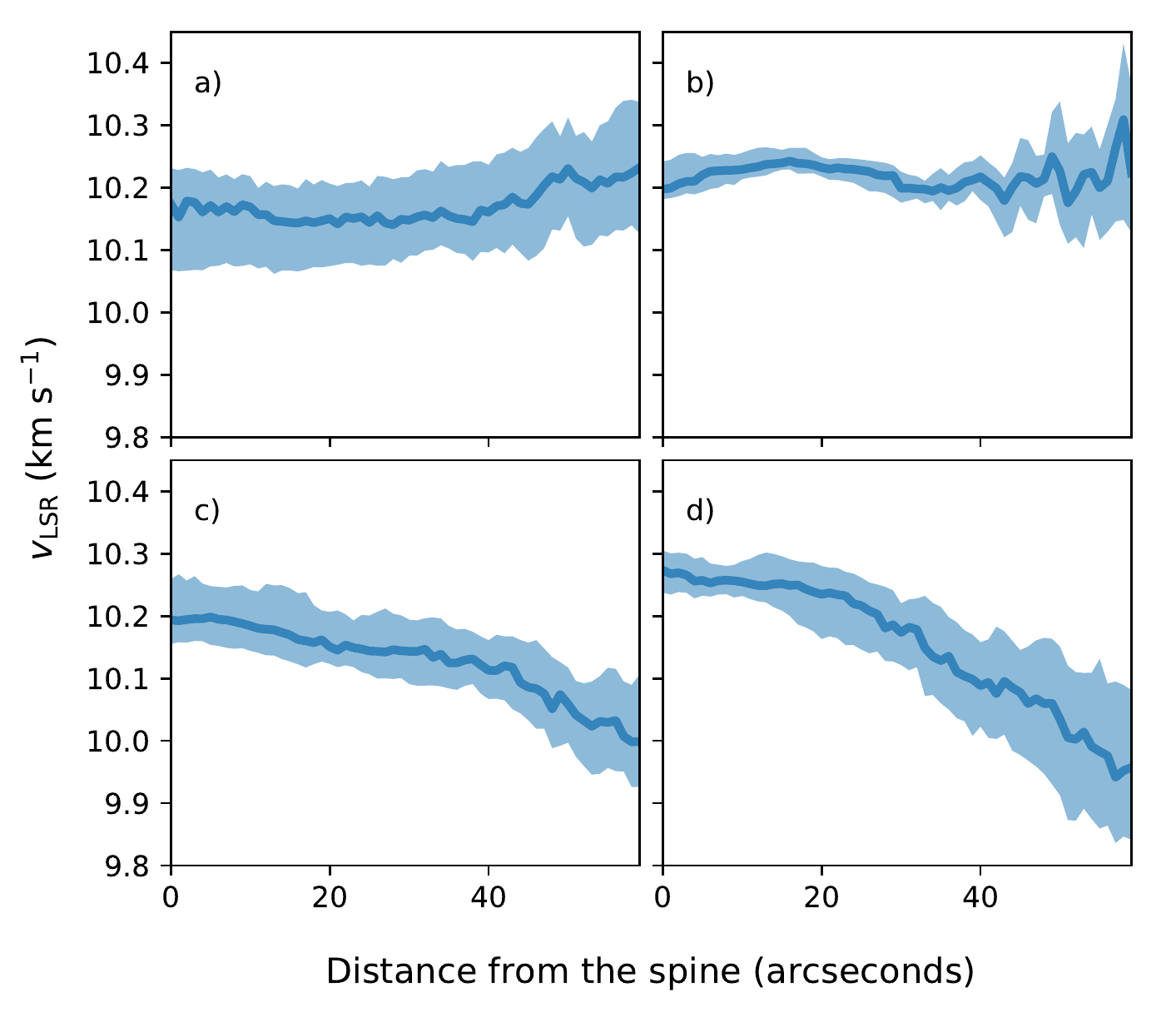}
\caption{Averaged radial profiles of $v_{\mathrm{LSR}}$ for the four B5 filaments. The solid lines show the median value of the profiles while the shaded regions represent the 25- and 75-percentile ranges of the profile.\label{fig:radPro_vlsr}}
\end{figure}

Figure \ref{fig:radPro_vlsr} shows the $v_{\mathrm{LSR}}$ radial profiles of the four B5 filaments. While $v_{\mathrm{LSR}}$ values of the southern filaments, i.e., \textit{a} and \textit{b}, remain relatively constant at all projected radii, the $v_{\mathrm{LSR}}$ values of the northern filaments, i.e., \textit{c} and \textit{d}, show clear downward trends at larger radii towards lower velocities. An examination of our final $v_{\mathrm{LSR}}$ map shown in Figure \ref{fig:spine_maps} reveals that most of these lower $v_{\mathrm{LSR}}$ pixels reside on the western and northern outskirts of B5, well outside the central subsonic region. While these transitions towards lower $v_{\mathrm{LSR}}$ values are rather gradual, the $v_{\mathrm{LSR}}$ change over the $60''$ scale (i.e., 0.09 pc) is comparable or greater than the sound speed of a 10 K gas ($\sim 0.2$ km s$^{-1}$).

\begin{figure}
\centering
\includegraphics[width=0.45\textwidth]{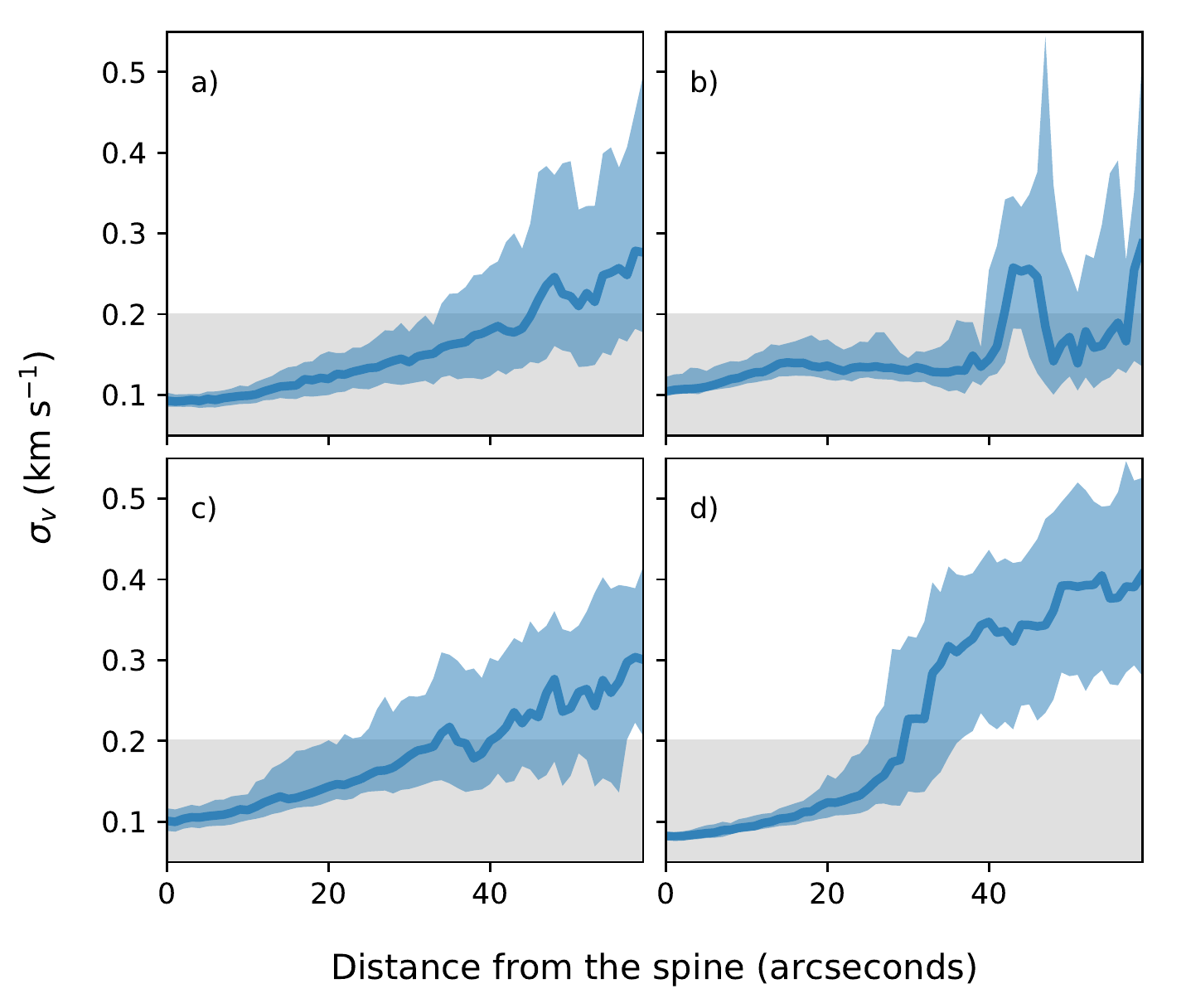}
\caption{Radial profiles of $\sigma_v$ for the four B5 filaments presented in the same way as Figure \ref{fig:radPro_vlsr}. The grey shaded region shown where the $\sigma_v$ should be if its none-thermal component is subsonic. \label{fig:radPro_sigv}}
\end{figure}

Figure \ref{fig:radPro_sigv} shows the radial $\sigma_v$ profiles of the B5 filaments. They tend to increase monotonically with radii and typically become supersonic at radii $>40''$, i.e., $\sim 0.06$ pc. For filaments \textit{c} and \textit{d}, these radii are also where the $v_{\mathrm{LSR}}$ values start to deviate from the ridge center by $\sim 0.1 - 0.2$ km s$^{-1}$, which are subsonic and transonic values, respectively. Incidentally, the 0.06 pc radius is also about where the background emission starts to dominate in these filaments according to our best-fit Plummer-like profiles (Eq. \ref{eq:Plummer}). This transition can be seen in Figure \ref{fig:radPro} where the grey shaded regions plotted over the integrated intensity profiles represent the background emission derived from our best-fit Plummer-like model. We note that filament \textit{b} is fit best with no background.

The fact the $\sigma_v$ measured in B5 tends to be supersonic where the background emission dominates over the filaments suggests the sharp increase of $\sigma_v$ seen towards filament \textit{d} around $r=30''$ in Figure \ref{fig:radPro_sigv} is not a physical feature inherent to that filament. The apparent $\sigma_v$ transition we see, which occurs at regions where only a single component is detected, likely results from the ambient emission being picked up by the spectral fitting routine as it starts to dominate. Such a transition at where ambient emission dominates over their dense, subsonic counterpart is also likely responsible for the sharp sonic transitions reported by lower spatial resolution NH$_3$ studies conducted with single component fits (e.g., \citetalias{Pineda2010}; \citealt{ChenHope2019}; \citealt{ChenHope2020}).

The $\sigma_v$ values found within two half widths at half maximum (HWHM) of the B5 filaments (i.e., $r \sim 20'' = 0.03$ pc) tend to increase linearly. The only exception is found in filament \textit{b}, where the profile appears fairly flat in the radial range of $15''-40''$. These flat or increasing trends appear consistent with those observed in subsonic cores in Taurus B18 and Ophiuchus L1688 by \cite{ChenHope2019}, which are marginally resolved at a $0.03$ pc scale by the GBT beam ($\sim 0.02$ pc in those clouds) and fitted with single-component models. 

In hydrodynamic simulations, the fiducial model of \cite{Heigl2020} also yields radial $\sigma_v$ profiles that increase linearly with radius throughout various time steps of the simulation. The maximum FWHM of their filament throughout its evolution is $\sim 0.03$ pc in the fiducial model, which is similar to those we measure in B5 and differs significantly from the typical FWHM $\sim 0.1$ pc value widely reported by dust-continuum studies (e.g., \citealt{Arzoumanian2011}). If this model indeed describes the actual filaments we observe well, then the linear correlation between $\sigma_v$ and the projected radii observed in B5 may be attributed to accretion-driven turbulence. Under the model by \citeauthor{Heigl2020}, such an anti-correlation between turbulence and gas density cannot produce a radial turbulent pressure gradient, and, consequently, the turbulence cannot provide additional radial support against self-gravity. 

Similarly, large volume simulations of filaments that include self-gravity, turbulence, magnetic fields, and outflow feedback also show line-of-sight $\sigma_v$ profiles that decrease towards the filament spines (\citealt{Federrath2016}; see their Figure 5). With turbulence primarily injected from the largest scales, \citeauthor{Federrath2016} attributes the top-down turbulent energy cascade as the reason behind this trend seen in their simulations, both in the sonic and subsonic regimes. Unlike those found in simulations by \cite{Heigl2020}, however, the $\sim 0.1$ pc filament widths measured in these simulations are about a factor of three larger than those we measure in B5.

Alternatively, if the radial inward motion of a filament takes the form of $v\left ( r \right ) \propto -r$ that describes the flat density inner regions of a gravitationally contracting, prestellar core (e.g., \citealt{Whitworth1985}), then the linear correlation between $\sigma_v$ and the projected radii we observe may be due to infall motions, rather than turbulence. Indeed, \cite{Vazquez-Semadeni2019} proposed such an interpretation under the GHC framework for similar $\sigma_v$ profiles seen towards subsonic dense cores (e.g., \citealt{ChenHope2019}). If the cloud turbulence is only mildly supersonic, as proposed in the GHC framework, such infall motions will dominate over their turbulent counterparts. Since density profiles of filaments and prestellar cores are both well described by Plummer-like functions, with $r < R_{\mathrm{flat}}$ corresponding to the flat inner region (e.g., \citealt{Whitworth2001}), an extrapolation of such an interpretation from cores to filaments is reasonable. The smooth $v_{\mathrm{LSR}}$ transition between the filaments and their ambient gas seen in Figure \ref{fig:radPro_vlsr} can be further understood as the smooth, non-shocked accretion flow predicted by the GHC.

To better test such an interpretation, we will need synthetic observations of simulations with resolution comparable to our data (i.e., $0.009$ pc). Specifically, such a study is required to confirm whether or not the infall motions proposed by \cite{Vazquez-Semadeni2019} for the GHC can remain spectrally unresolved along sightlines to produce radial $\sigma_v$ profiles similar to those shown in Figure \ref{fig:radPro_sigv}. Synthetic observations and follow-up observations will also be needed to distinguish the predictions of the GHC models from that of \cite{Heigl2020}, where the latter expects accretion shocks. In particular, such studies will need to determine whether or not the gradual $v_{\mathrm{LSR}}$ profiles seen in Figure \ref{fig:radPro_vlsr} can arise from projection and chemical effects even when a shock exists between a filament and the accretion flow.

Despite the presence of an outflow in B5 driven by B5-IRS1 (e.g., \citealt{Stephens2019S}), we found no combined spatial and spectral correlation between the outflows and the high $\sigma_v$ regions. This result suggests the slightly higher $\sigma_v$ values seen towards the `valley' between filament \textit{a} and \textit{b} are not driven by the IRS1 outflow. Instead, such high $\sigma_{v}$ regions may be associated with ongoing infall towards the B5-IRS1 protostar.

\subsection{Velocity Gradients \label{subsec:vgrads}}

\begin{figure*}[ht!]
\centering
\includegraphics[width=0.975\textwidth]{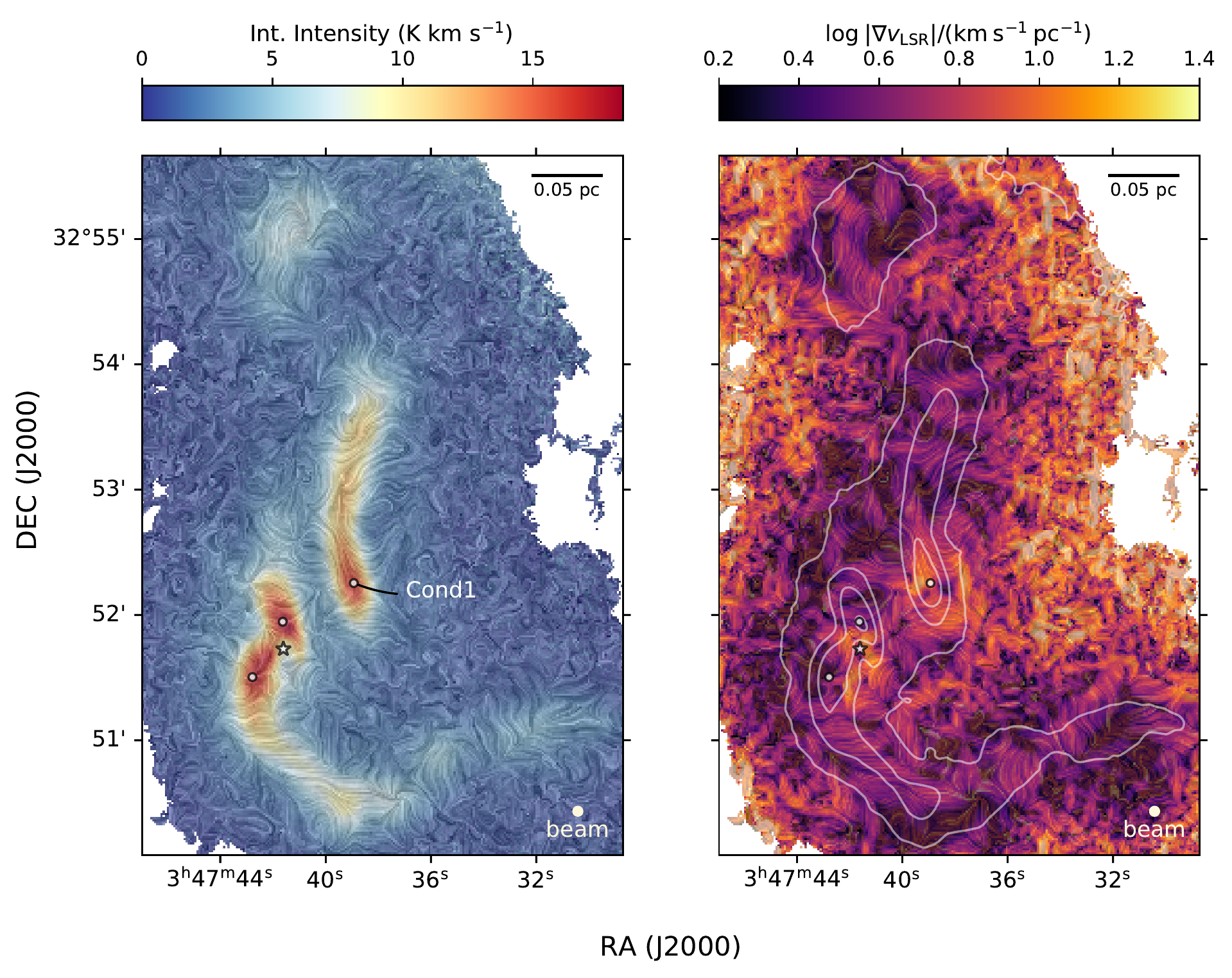}
\caption{Line integral convolution of the $\nabla v_{\mathrm{LSR}}$ field overlaid on the integrated intensity map derived from our NH$_3$ fits (left) and the $\log{|\nabla v_{\mathrm{LSR}}|}$ map (right). The white contours show where the modelled integrated intensity are at 2.5 K km s$^{-1}$, 7.5 K km s$^{-1}$, 12.5 K km s$^{-1}$, and 17.5 K km s$^{-1}$, respectively. The position of the IRS1 protostar and condensations identified by \citetalias{Pineda2015} are marked by the star and circles, respectively. The beam of the data and the physical scale bar are shown in both panels. \label{fig:lic_maps}}
\end{figure*}

Following \citetalias{ChenM2020}, we calculate the velocity gradient, $\nabla v_{\mathrm{LSR}}$, of our final, velocity-coherent $v_{\mathrm{LSR}}$ map at each pixel by least-square fitting a plane over the $v_{\mathrm{LSR}}$ values found within a circular aperture centered on the pixel. We set the diameter of the aperture to be 10 pixels wide, about twice the beam size, to ensure we measure gradients over well-resolved structures. The resulting $\nabla v_{\mathrm{LSR}}$ calculated for B5 reveals a highly complex field that is not expected from either a body of subsonic gas or smooth accretion flows. To well capture such complexity without reducing it to simple statistics, we focus much of our $\nabla v_{\mathrm{LSR}}$ discussion here on descriptive analyses.

Figure \ref{fig:lic_maps} shows the resulting $\nabla v_{\mathrm{LSR}}$ visualized with line integral convolution (LIC; \citealt{Cabral1993}) textures overlaid on top of the NH$_3$ integrated intensity map (left) and the $\log{|\nabla v_{\mathrm{LSR}}|}$ map (right) of our observations. We use the \texttt{LicPy} software for our LIC visualizations \citep{Rufat2018}. Interestingly, the local $\nabla v_{\mathrm{LSR}}$ field in B5 tends to be well ordered in the brightest (i.e., densest) parts of the four filaments and becomes increasingly less organized further away from them, likely due to the increasing contribution of noise. In well-ordered regions, the $\nabla v_{\mathrm{LSR}}$ orientations tend to be similar on the $\sim 30''$ (i.e. $\sim 0.04$ pc) scale in and around the filaments and remain so throughout the lengths of the shorter filaments (i.e., \textit{b} and \textit{d}), but not their longer counterparts (i.e., \textit{a} and \textit{c}). The southeastern segment of filament \textit{a}, for example, prominently features a well-ordered $\nabla v_{\mathrm{LSR}}$ that runs across the filament at about a $45\degree$ angle relative to the spine.

The $\nabla v_{\mathrm{LSR}}$ orientations at the fainter, western end of filament \textit{a} and the brighter southern end of filament \textit{c} also seem to vary as they move towards the filament spine. Specifically, the $\nabla v_{\mathrm{LSR}}$ at these locations tend to run perpendicularly to the filaments at larger radii and become parallel as they get closer to the filament spines. While the observed $\nabla v_{\mathrm{LSR}}$ may not necessarily map well onto the underlying, three-dimensional velocity fields, these radial trends do qualitatively resemble the velocity fields seen in simulations by \cite{Gomez2014}. In their simulations, which do not account for magnetic fields, gas tends to fall perpendicularly onto the filament and then starts to flow in parallel with the spine. We note, however, that the morphology of such a velocity field in projection may not necessarily imprint itself well onto the observed $\nabla v_{\mathrm{LSR}}$, which measures the acceleration or deceleration of line-of-sight velocities. 

The right panel in Figure \ref{fig:lic_maps} shows the LIC visualization of $\nabla v_{\mathrm{LSR}}$ plotted over its log magnitude ($\log{|\nabla v_{\mathrm{LSR}}|}$) map. Interestingly, the locations of B5 closer to filaments tend to contain `bands' of high $|\nabla v_{\mathrm{LSR}}|$ structures that are often $\sim 10''-20''$ in width and $\sim 60''-120''$ in length. Furthermore, the $\nabla v_{\mathrm{LSR}}$ is very ordered within these high $|\nabla v_{\mathrm{LSR}}|$ bands, usually running along the same direction within each band. The locations of these band-like $|\nabla v_{\mathrm{LSR}}|$ structures do not correlate well with the bright (i.e., high column density) structures seen in the integrated intensity map, except for filament \textit{a}.

\begin{figure*}[ht!]
\centering
\includegraphics[width=0.95\textwidth]{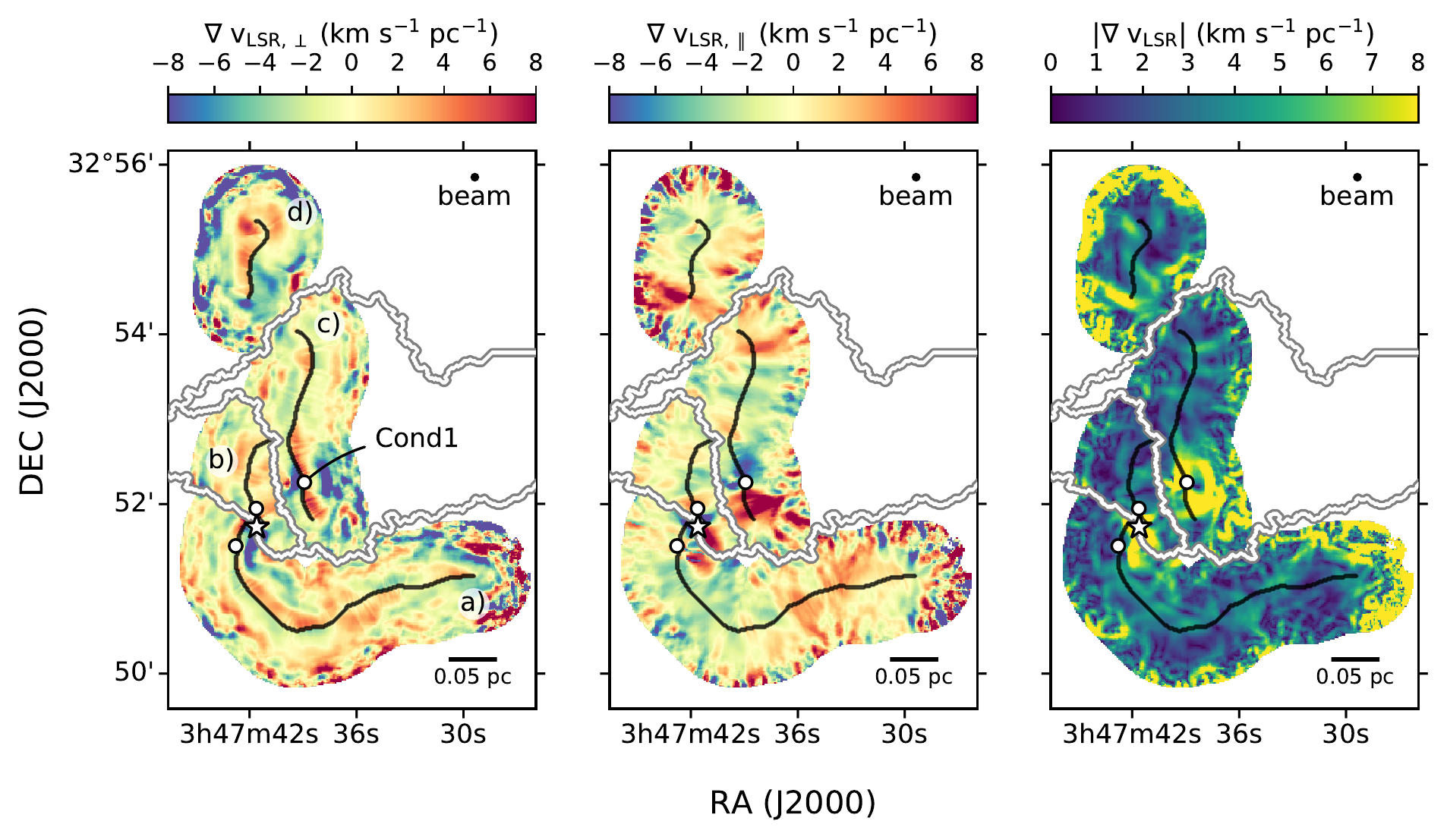}
\caption{The spatial distributions of the perpendicular and parallel components of $\nabla v_{\mathrm{LSR}}$ in B5, measured relative to the four filament spines (black), shown in the left and center panels, respectively. The decomposition of the $\nabla v_{\mathrm{LSR}}$ into its two components is performed within the watershed boundaries (white) of their associated spines. The spatial distribution of the total $\nabla v_{\mathrm{LSR}}$ magnitude is shown in the right panel. The position of the IRS1 protostar and condensations identified by \citetalias{Pineda2015} are marked by the star and circles, respectively. The beam of the data and a physical scale bar are shown in all panels. \label{fig:vgrad_comp_maps}}
\end{figure*}

Figure \ref{fig:vgrad_comp_maps} shows the $\nabla v_{\mathrm{LSR}}$ measured in B5 decomposed into components that are perpendicular (i.e., $\nabla v_{\mathrm{LSR, \perp}}$; left panel) and parallel ($\nabla v_{\mathrm{LSR, \parallel}}$; center panel) to the filament spines (black lines) within their respective watershed boundaries (white lines). We accomplished this decomposition using the technique developed by \cite{ChenM2020}. The total magnitudes of the non-decomposed $\nabla v_{\mathrm{LSR}}$ are shown in the right panel. Following the convention set by \cite{ChenM2020}, positive $\nabla v_{\mathrm{LSR, \perp}}$ values point away from the filament spines and vice versa for the negative values. By the same convention, positive $\nabla v_{\mathrm{LSR, \parallel}}$ values point away from the end of the filament furthest away from the image origin, i.e., the southeastern corner, and vice versa for negative values. Pixels more than $40''$ away from the filament spines are masked out due to noise concerns and excluded from our remaining analyses.

The $\nabla v_{\mathrm{LSR}}$ maps of B5 in Figure \ref{fig:vgrad_comp_maps} contain a wealth of small-scale, beam-resolved structures within $\sim 40''$ of the filament spine. The typical values of these compact structures in both the perpendicular and parallel directions are in the range of $1 - 5$ km s$^{-1}$ pc$^{-1}$. Even though $\nabla v_{\mathrm{LSR}}$ at pixels beyond $\sim 40''$ is also rich in structures (see the right panel of Figure \ref{fig:lic_maps}), the typical widths of these structures and the typical distances between them appear to be around the same size as our $6''$ beam. Examining the B5 $\nabla v_{\mathrm{LSR}}$ field with LIC (see Figure \ref{fig:lic_maps}) further reveals that the $\nabla v_{\mathrm{LSR}}$ orientations in these regions are highly disordered. For these reasons, we do not include pixels more than $40''$ away from the filament spines in our analysis and have masked out these locations accordingly.

Within $40''$ of filament spines, we find no $v_{\mathrm{LSR}}$ discontinuities between any of the filaments in B5. As a result, these filaments likely belong to the same parental, subsonic structure. The lack of $v_{\mathrm{LSR}}$ discontinuity between filaments also indicates that our kinematic analysis does not depend sensitively on the location of the watershed-defined borders that delineate the faint inter-filament emission. Moreover, this lack of $v_{\mathrm{LSR}}$ discontinuity also contrasts greatly with the kinematic behaviour of sub-filamentary fibers observed by \cite{Hacar2013} in Taurus with C$^{18}$O, which have supersonic levels of $v_{\mathrm{LSR}}$ differences ($\gtrsim 0.4$ km s$^{-1}$) between them. Similarly, fibre-like `sub-filaments' seen in simulations by \cite{Clarke2017} also have supersonic levels of line-of-sight velocity differences ($\sim 0.5 - 2$ km s$^{-1}$). The filaments we find in B5 may thus represent objects that are either evolutionarily or intrinsically different from fibre-like features found in both observation and simulations.

\subsubsection{Perpendicular velocity gradient \label{subsub:vgrad_perp}}

The $\nabla v_{\mathrm{LSR, \perp}}$ values measured in B5 tend to form elongated structures near the filament spines, often running parallel to the spines (see Figure \ref{fig:vgrad_comp_maps}). These $\nabla v_{\mathrm{LSR, \perp}}$ structures typically have magnitudes of $5-8$ km s$^{-1}$ pc$^{-1}$ and have opposite signs on the two sides of the spines, which indicates that the $\nabla v_{\mathrm{LSR, \perp}}$ field at those locations runs through, i.e., across, their respective filament spines rather than towards or away from the spines, as per the convention. The widths of these structures are typically about $8-12''$ and the $|\nabla v_{\mathrm{LSR, \perp}}|$ values of these structures are relatively constant throughout the structures. 

If we assume a planar-like accretion geometry, such as that illustrated by \citeauthor{Dhabal2018} (\citeyear{Dhabal2018}; see their Figure 15), and a $v(r) \propto -r$, prestellar-core-like, infall profile (e.g., \citealt{Whitworth1985}), then we should expect a constant velocity gradient across the filament spine, i.e., $\nabla v_{\mathrm{LSR, \perp}} \propto - \textrm{constant}$. Specifically, we should see such a profile within the flat-density, inner regions of the filament, i.e., at radii where the assumed infall would hold for prestellar cores. Indeed, the typical $\sim 10''$ widths of $\nabla v_{\mathrm{LSR, \perp}}$ structures we found near the spines are comparable to the $R_{\mathrm{flat}}$ values we derived from the best-fit Plummer-like filament profiles ($\sim 6-11''$), suggesting that these observed $\nabla v_{\mathrm{LSR, \perp}}$ are signs of mass infall. Considering that planar-like accretion flows are commonly found in theoretical models, ranging from sheets fragmenting into filaments (e.g., \citealt{Miyama1987}) to post-shock accretion flows along magnetic fields (e.g., \citealt{ChenCheYu2014}, \citealt{ChenCheYu2015}), a planar-like geometry for such infall is plausible.

Interestingly, elongated, high $\nabla v_{\mathrm{LSR, \perp}}$ structures are not always symmetrically found on both sides of spines. For example, the eastern half of filament \textit{a} exhibits such a structure only on its northwestern side. Furthermore, these elongated, high $\nabla v_{\mathrm{LSR, \perp}}$ structures are not always found near spines. Several are found at larger radii, i.e., $r \gtrsim 10''$, often running parallel to those adjacent to the spine but with an opposite sign. Such `secondary' $\nabla v_{\mathrm{LSR, \perp}}$ structures can be found in the eastern segment of filament \textit{a} and the southern segment of \textit{c}. While the occurrence of these elongated, sign-alternating $\nabla v_{\mathrm{LSR, \perp}}$ structures may seem peculiar at first, such a behaviour could correspond to compression waves similar to those found in models of \cite{Whitworth1985}.

Alternatively, similar line-of-sight velocity profiles can also be found in numerical simulations of accreting filaments by \cite{Clarke2017}. Specifically, the velocity structures in these simulations should produce $\nabla v_{\mathrm{LSR, \perp}}$ structures that are also elongated and relatively parallel to the spine. Despite having fairly radially symmetric initial conditions, the implied $\nabla v_{\mathrm{LSR, \perp}}$ resulted from the simulation is often asymmetrically distributed about the filament spines with radially-alternating signs, much like what we see in B5. Rather than compression waves, these elongated $\nabla v_{\mathrm{LSR, \perp}}$ structures arise from gas vorticity, i.e., $\mathbf{\omega} \equiv \nabla \times \mathbf{v}$, that runs predominantly parallel to filament spines and is likely driven by the radial accretion of an inhomogeneous (e.g., clumpy) flow onto the filament. If such vorticity is indeed responsible for the $\nabla v_{\mathrm{LSR, \perp}}$ observed in B5, then the partial, parallel overlap between filament \textit{b} and \textit{c} may have been produced from a process similar to the ``fray'' step of the ``fray and fragment'' fiber formation scenario \citep{Tafalla2015}. 

The $v_{\mathrm{LSR}}$ differences between the individual B5 filaments ($\lesssim 0.2$ km s$^{-1}$) are much smaller than those expected in the ``fray and fragment'' scenario. Indeed, the line-of-sight velocity differences between fibers found in observations (e.g., \citealt{Hacar2013}) and the sub-filaments found in the simulations (e.g., \citealt{Clarke2017}) are mostly supersonic. Further theoretical work is thus needed to determine if such discrepancies can be reconciled via filament properties such as mass, age, or the initial turbulence level. For example, the simulations by \cite{Clarke2017} involve initial conditions resembling filaments that are generally more massive than those found in our B5 observations, calibrated empirically with the works of \cite{KirkHelen2013} and \cite{Palmeirim2013}. Having initial conditions that better represent the B5 filaments in models are therefore needed to resolve whether or not the apparent tension between the $v_{\mathrm{LSR}}$ differences observed in B5 and the earlier simulations by \cite{Clarke2017} can be resolved.

\subsubsection{Parallel velocity gradient \label{subsub:vgrad_para}}

The $\nabla v_{\mathrm{LSR, \parallel}}$ map shown in Figure \ref{fig:vgrad_comp_maps} (center) also displays a wealth of small-scale structures. The morphology of these structures, however, is less well defined than their $\nabla v_{\mathrm{LSR, \perp}}$ counterparts in general. The typical magnitudes of these $\nabla v_{\mathrm{LSR, \parallel}}$ structures are about $3-6$ km s$^{-1}$ pc$^{-1}$.

Two prominent, opposite-signed $\nabla v_{\mathrm{LSR, \parallel}}$ structures can be seen towards the southern end of filament \textit{c}. These structures have the highest $\nabla v_{\mathrm{LSR, \parallel}}$ magnitudes found in B5 (i.e., $5-12$ km s$^{-1}$ pc$^{-1}$) and are located adjacent to each other, centered around the \textit{B5-Cond1} `condensation' identified by \citetalias{Pineda2015} from these same data. The boundary between these two adjacent $\nabla v_{\mathrm{LSR, \parallel}}$ structures is spatially correlated with \textit{B5-Cond1}'s emission peak, indicating the $v_{\mathrm{LSR}}$ values measured in filament \textit{c} are locally increasing towards the emission peak from both sides. Examining the $v_{\mathrm{LSR}}$ map around this location (see Figure \ref{fig:spine_maps}) confirms that the $v_{\mathrm{LSR}}$ values in the southern half of filament \textit{c} are indeed higher in \textit{B5-Cond1} than in its surroundings, by as much as 0.15 km s$^{-1}$.

While a spatial association between $\nabla v_{\mathrm{LSR, \parallel}}$ and a density peak may indicate gas flows toward a dense core along the filament axis, the sign flip we observe at the emission peak is unexpected in such an interpretation. For example, if we model a filament as a cylinder inclined towards the observer (e.g., Figure 16 by \citealt{Dhabal2018}), then infall flow towards a dense core with a $v\left ( r \right ) \propto -r$ profile (e.g., \citealt{Whitworth1985}) along the filament axis would produce a constant $\nabla v_{\mathrm{LSR, \parallel}}$ across the core without a sign change. In fact, any infall velocity profile that increases monotonically away from the center of a dense core should not produce a sign change in $\nabla v_{\mathrm{LSR, \parallel}}$ across the center of the core. 

We speculate the sign change in $\nabla v_{\mathrm{LSR, \parallel}}$ across \textit{B5-Cond1} indicates that the southern end of filament \textit{c} is curved towards our line of sight in three-dimensions. Considering that all four B5 filaments are somewhat curved on the 0.1 pc dense core scale ($\sim 60''$), the assumption these filaments are straight on this scale may indeed be poor. In this case, the observed $\nabla v_{\mathrm{LSR, \parallel}}$ may correspond to longitudinal infall towards \textit{B5-Cond1} along a filament that is curved and orientated such that the two ends of \textit{B5-Cond1} are inclined either both towards or away from the observer.

Indeed, \cite{Smith2016} showed simulations with line-of-sight (LOS) velocity profiles along filaments that can produce $\nabla v_{\mathrm{LSR, \parallel}}$ features that change signs across the center of a core. For example, two out of four cores along a series of connected sub-filaments in their \textit{S1F1T303} snapshot show a `U' shape LOS velocity profile that would produce such a sign change. The LOS velocity profiles of the other two cores, however, appeared fairly linear, which is consistent with the velocities expected from a relatively straight filament with a $v\left ( r \right ) \propto -r$ infall profile.

On larger scales, filaments simulated by \cite{Gomez2014} also showed velocity profiles that may produce $\nabla v_{\mathrm{LSR, \parallel}}$ sign changes across several cores. While LOS velocities were not directly presented for these simulations, the filaments in these simulations do contain sharp bends at the locations of dense cores as well. These results further suggest that a sign change around $\nabla v_{\mathrm{LSR, \parallel}}$ toward a dense core may indicate a filament bent along the lines of sight.

Similar to the findings of \cite{ChenM2020} in Perseus NGC 1333 with a lower resolution ($\sim 0.05$ pc) study, filaments in B5 also seem to contain ``zebra-strip-like'' $\nabla v_{\mathrm{LSR, \parallel}}$ structures, particularly in filament \textit{c}. Due to their lack of correlation with local overdensities (i.e., condensations), except for \textit{B5-Cond1}, these zebra stripes may indicate the presence of magnetohydrodynamic (MHD) waves like those proposed by \cite{Heyer2016} to explain the lower-density striations seen in the Taurus molecular cloud. As proposed by \cite{Offner2018}, these waves can be excited by protostellar outflows. While we found no evidence that B5-IRS1's outflows significantly impact the ammonia-traced dense gas directly (also see \ref{subsub:radial_vel}), the presence of these zebra stripes may suggest the outflows' energy is being transferred onto the denser structures still via MHD waves.

\subsubsection{Global trends in velocity gradients \label{subsub:vgrad_global}}

\begin{figure}[t!]
\centering
\includegraphics[width=0.45\textwidth]{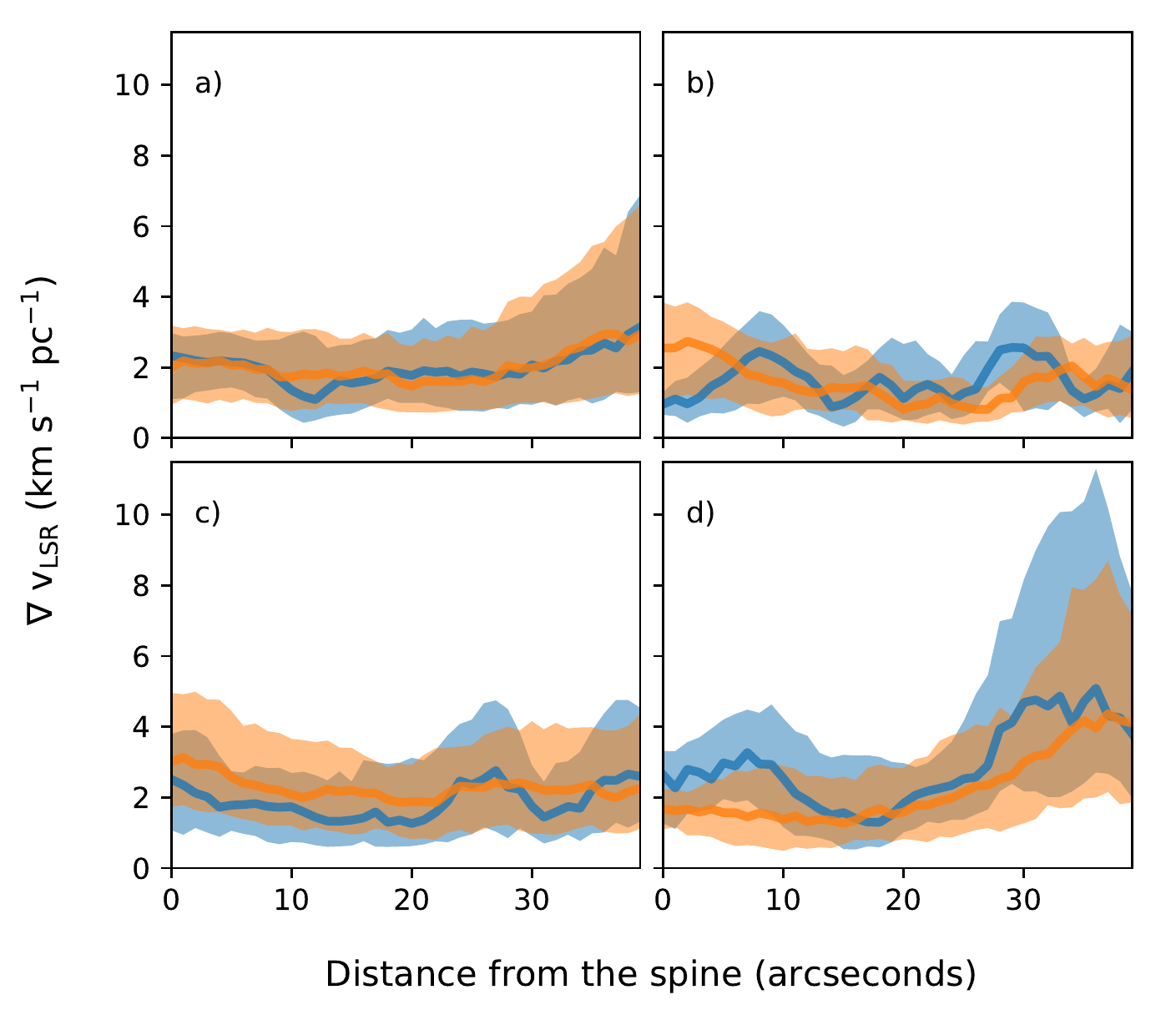}
\caption{The radial profiles of $| \nabla v_{\mathrm{LSR, \perp}} |$ (blue) and $| \nabla v_{\mathrm{LSR, \parallel}} |$ (orange) of the four filaments identified in B5 with the solid lines showing the median values while the shaded regions show the 25- to 75-percentile ranges.  \label{fig:radPro_vgrad_comps}}
\end{figure}

Figure \ref{fig:radPro_vgrad_comps} shows radial profiles of $| \nabla v_{\mathrm{LSR, \perp}} |$ (blue) and $| \nabla v_{\mathrm{LSR, \parallel}} |$ (orange) of the four B5 filaments. On scales larger than $\sim 10''$, the radial trends of $| \nabla v_{\mathrm{LSR, \perp}} |$ and $| \nabla v_{\mathrm{LSR, \parallel}} |$ effectively have the same shape. This result is different from that seen by \cite{ChenM2020} on larger scales ($\sim 0.05$ pc) in Perseus NGC 1333, where $| \nabla v_{\mathrm{LSR, \perp}} |$ and $| \nabla v_{\mathrm{LSR, \parallel}} |$ can have distinctly different profiles. If the velocity gradients measured in B5 arise from acceleration or deceleration associated with accretion flows, then such a trend suggests that the radial (i.e., perpendicular) and longitudinal (i.e., parallel) accretion flows share the same behaviour regardless of their distance from a filament spine.

In addition to being similar to each other, the radial profiles of $| \nabla v_{\mathrm{LSR, \perp}} |$ and $| \nabla v_{\mathrm{LSR, \parallel}} |$ in the B5 filaments are also fairly constant, i.e., flat. The only exception are the profiles found in filament \textit{d}. The median values of these constant profiles are in the range of $1.7 - 2.3$ km s$^{-1}$ pc$^{-1}$, which are within the range of large-scale ($> 0.2$ pc) velocity gradients measured from various filaments in the Gould Belt Clouds ($\sim0.5 - 5$ km s$^{-1}$ pc$^{-1}$; e.g., \citealt{KirkHelen2013}, \citealt{Fernandez-Lopez2014}, \citealt{LeeKatherine2014}, \citealt{Hacar2017}).

Naively, the flat radial profiles of $| \nabla v_{\mathrm{LSR, \perp}} |$ measured in B5 filaments outside of the $R_{\mathrm{flat}}$ region seem contrary to those expected from gravitationally dominated accretion flows, whose velocity gradient magnitudes are expected to increase towards the filament spine. Indeed, free-fall accretion towards an infinitely long cylinder would have a radial velocity profile of $v_{\mathrm{ff}}(r) \propto [ \ln(r_{0}/r) ]^{1/2}$ (e.g., \citealt{Heitsch2009}) that produces gradients in the form of $dv_{\mathrm{ff}}/dr \propto [r^2 \ln(r_{0}/r) ]^{-1/2}$, which increase monotonically towards the filament spine for $r$ less than about half of where the accretion flow was initially at rest (i.e., $r_{0}$). Similarly, the average radial velocity profile taken from simulations by \cite{Gomez2014} (see their Figure 10) has a roughly log-linear profile, i.e., $v_{\perp}(r) \propto \log_{10}(r)$, for $r$ in the range of 0.01 pc to 0.08 pc, which also produces gradients that increase monotonically towards the filament spines in the form of $dv_{\perp}/dr \propto r^{-1}$. 

We note that further projection modeling and abundance matching are needed to properly compare these theoretical velocity gradients to those we observed here. Furthermore, if the averaged $| \nabla v_{\mathrm{LSR, \perp}} |$ radial profiles in B5 are dominated by small-scale velocity fluctuations, such as those driven by macroscopic turbulence, instead of accretion flows, then the $| \nabla v_{\mathrm{LSR, \perp}} |$ cannot be used as a probe of accretion signatures. Future work that investigates the nature of these measured $| \nabla v_{\mathrm{LSR, \perp}} |$ is needed to interpret these results further. 

\begin{figure}[t!]
\centering
\includegraphics[width=0.48\textwidth, trim={7mm 10mm 0 4mm}]{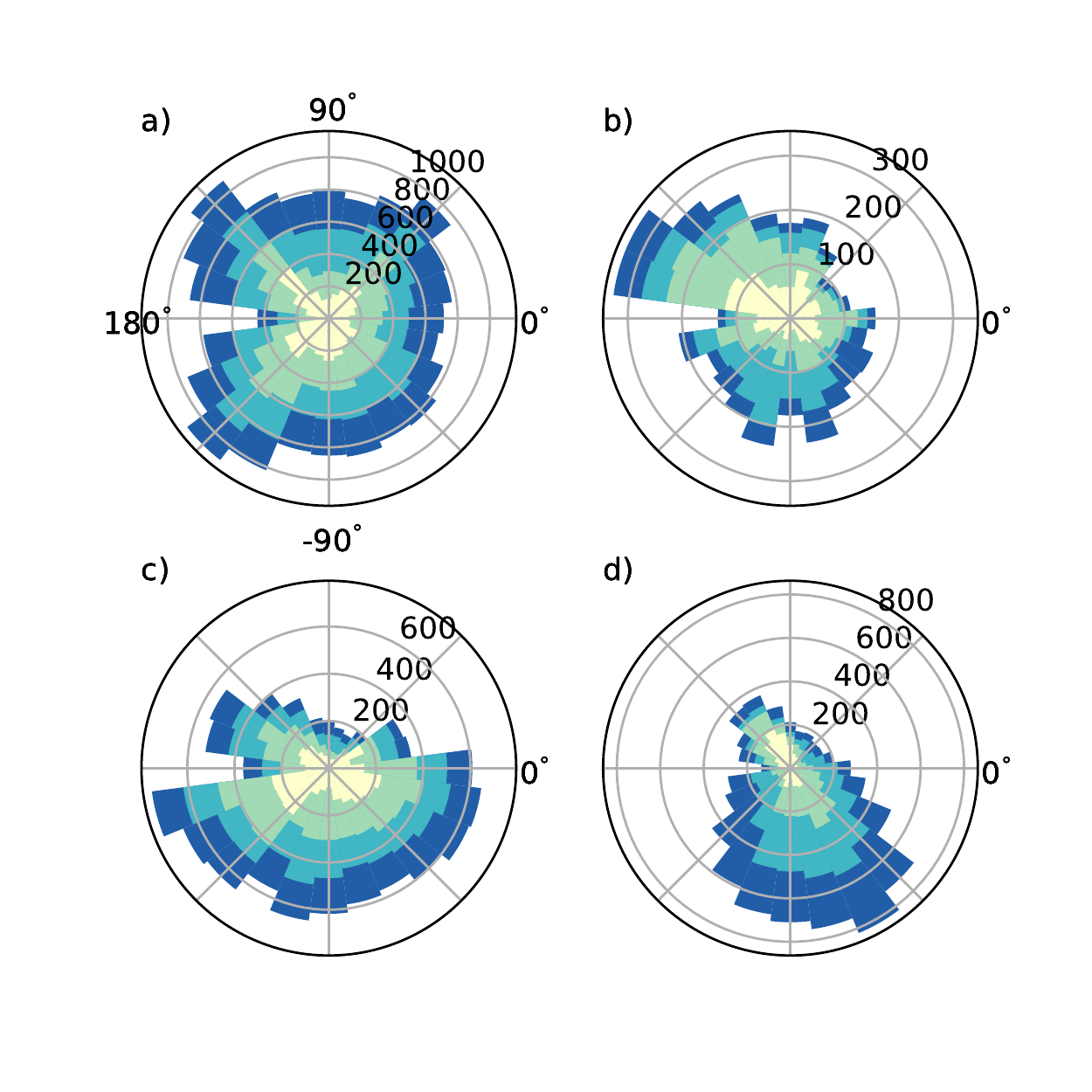}
\caption{Polar histograms of $\nabla v_{\mathrm{LSR}}$ orientations measured in the B5 filaments. By convention, $0\degree$ points along the filament spine, away from the end closest to the origin of the image, and $90\degree$ and $-90\degree$ point away and towards the spine, respectively. The light to dark colors correspond to contributions from pixels in the radial bins from $r= 0 - 40''$, divided into increments of $10''$. Pixels with estimated $\nabla v_{\mathrm{LSR}}$ errors larger than their magnitudes are excluded from the sample.\label{fig:vgrad_polar_hist}}
\end{figure}

Figure \ref{fig:vgrad_polar_hist} shows polar histograms of the $\nabla v_{\mathrm{LSR}}$ orientation measured in the B5 filaments, with radial axes showing the number of pixels in each angular bin. By convention, the $0\degree$ angle points along the filament spine, away from the end closest to the origin of the image. The $90\degree$ and $-90\degree$ angles, on the other hand, point away and towards the spine, respectively. The light to dark colors correspond to contributions from pixels in the radial distance bins from $r= 0'' - 40''$, divided into increments of $10''$. Pixels with estimated $\nabla v_{\mathrm{LSR}}$ errors larger than their magnitudes are excluded from the sample. 

With the exception of filament \textit{a}, the $\nabla v_{\mathrm{LSR}}$ of the B5 filaments is not randomly distributed, consistent with what \cite{ChenM2020} found for filaments in Perseus NGC 1333 at the $\sim 0.05$ pc scale. The orientations found in B5 on the $\sim 0.009$ pc scale, however, are much less unimodally or bimodally distributed than those in NGC 1333. Given that the small-scale $\nabla v_{\mathrm{LSR}}$ field we measure in B5 appears fairly ordered on the $\sim 0.05$ pc scale (see Figure \ref{fig:lic_maps}), observations made at a resolution comparable to such a scale will average-out the smaller-scale variations and produce more pronounced trends of $\nabla v_{\mathrm{LSR}}$ orientations, as seen in NGC 1333.

Figure \ref{fig:vgrad_polar_hist} also shows the $\nabla v_{\mathrm{LSR}}$ orientations measured in B5 tend to be less random for pixels located near the spines than those measured further away. Since these preferential orientations on small scales can differ significantly from region to region on scales $\gtrsim 30''$ (see Figure \ref{fig:lic_maps}), the increase in randomness at larger radii bins is thus due to the larger size of such bins encompassing a larger area of the sky than their smaller counterparts. While these radial trends do not necessarily imply that the $\nabla v_{\mathrm{LSR}}$ orientations are intrinsically more random further away from the filament spines, they do illustrate that the well-organized the small-scale $\nabla v_{\mathrm{LSR}}$ orientations measured in B5 ($\sim 10''$) do not remain the same on larger scales ($\gtrsim 80''$). This result suggests that highly directional fields, such as gravity or a magnetic field, do not impose a strong order in B5 on the 0.1 pc (i.e., $\sim 80''$) scale. Indeed, we do not see preferred parallel or perpendicular $\nabla v_{\mathrm{LSR}}$ field associated with the B5 filaments globally.

\section{Summary and Conclusions \label{subsec:conclusion}}

In this paper, we performed two-component spectral fits (where appropriate) to the $6''$ resolution, NH$_3$ (1,1), GBT+VLA combined observations of the Perseus B5 region, first published by \citetalias{Pineda2015}. We fit the observed spectra automatically using the \texttt{MUFASA} package (\citetalias{ChenM2020}) over two iterations: first with the spatially convolved data using moment-based initial guesses followed by a second, full-resolution fit using results from the first iteration as its initial guesses. The two iteration approach takes advantage of the SNR boost and an increase in spatial awareness that comes with the convolved data. Finally, \texttt{MUFASA} uses the AICc criteria (\citealt{Akaike1974}; \citealt{Sugiura1978}) to select the best spectral model between the noise, a one-component fit, and a two-component fit models. 

We identified filaments in ppv space from our best-fit model emission with the hyperfine structures removed. We accomplished such an identification using the \texttt{CRISPy} software (\citetalias{ChenM2020}). Given that the B5 region seems to be dominated by an overall kinematically coherent structure that is well traced by our one-component fits, which is also detected in regions where a second component is present, we adopted only one of the two-components that is kinematically most similar to one-component fits in neighbouring pixels into our final $v_{\mathrm{LSR}}$ and $\sigma_v$ maps. We subsequently computed the velocity gradient field, i.e., $\nabla v_{\mathrm{LSR}}$, from the resulting, final $v_{\mathrm{LSR}}$ map on the $10''$ scale ($\sim 2$ beam widths) and performed a detailed, high-resolution analysis of the filament kinematics on these final $v_{\mathrm{LSR}}$ and $\sigma_v$ maps.

The main results of our analyses are summarized as follows:

\begin{enumerate}
    \item The filamentary structures resolved by our data on the 0.009 pc ($6''$) scale not only have subsonic levels of internal $\sigma_v$ (see also \citetalias{Pineda2011}) but also have subsonic levels of $v_{\mathrm{LSR}}$ (i.e., bulk motions) relative to one another. The latter behavior contrasts greatly from those found in fibers observed by \cite{Hacar2011} with C$^{18}$O and in sub-filaments synthetically observed with C$^{18}$O in simulations by \cite{ClarkeS2018}, both of which have supersonic levels of relative LOS motions.
  
    \item The B5 filaments have averaged integrated intensity radial profiles that resemble the Plummer-like functions remarkably well. The best-fit Plummer-like functions to filaments \textit{a}, \textit{b}, \textit{c}, and \textit{d} have best-fit power-law index $p$ values of $2.5 \pm 0.1$, $2.3 \pm 0.1$, $3.1 \pm 0.1$, and $3.5 \pm 0.4$, respectively. Interestingly, the $p$ values found for filaments \textit{c} and \textit{d} are higher than those typically found in dust continuum studies ($\sim 2$; e.g., \citealt{Arzoumanian2011}).
    
    \item The FWHM widths of filaments \textit{a}, \textit{b}, \textit{c}, and \textit{d} inferred from best-fit Plummer-like functions are $0.025 \pm 0.01$ pc, $0.029 \pm 0.02$ pc, $0.023 \pm 0.01$ pc, and $0.028 \pm 0.03$ pc, respectively. These values are significantly smaller than those often found by dust continuum studies (e.g., \citealt{Arzoumanian2011}) but consistent with those found by dense gas tracer studies (e.g., N$_2$H$^+$ in Orion A; \citealt{Hacar2018}).
  
    \item The radial $\sigma_v$ profiles of the B5 filaments tend to increase monotonically and linearly towards larger radii within the subsonic regime. Such a behaviour is similar to that found by \cite{Heigl2020} in their isolated, simulated filaments with accretion-driven turbulence, as well as that found by \cite{Federrath2016} in his large volume simulation with supersonic-turbulence driven on the cloud scale. Alternatively, if the non-thermal component of the observed $\sigma_v$ corresponds to unresolved infall motion rather than turbulence, as proposed by \cite{Vazquez-Semadeni2019} for mildly supersonic clouds, then the radial $\sigma_v$ profiles we measured in the B5 filaments may correspond to $v\left ( r \right ) \propto -r$ type of prestellar infall motion (e.g., \citealt{Whitworth1985}).
    
    \item The $\nabla v_{\mathrm{LSR}}$ field we measured in B5 on the $10''$ scale appears to be well ordered locally, often on scales of $\sim 30''$ ($\sim 0.04$ pc), but not always consistently across scales of an entire filament or larger.
  
    \item The $\nabla v_{\mathrm{LSR}}$ component perpendicular to the filament spines (i.e., $\nabla v_{\mathrm{LSR}, \perp}$) often contains compact, elongated structures running parallel to the spines, with magnitudes typically $>4$ km s$^{-1}$ pc$^{-1}$. While structures within inner regions of the filaments may indicate $v\left ( r \right ) \propto -r$ type of infall, the alternating sign change of these structures outside the inner regions indicates additional physical processes at play. Such processes may include compression waves (e.g., \citealt{Whitworth1985}) or velocity vortices (i.e., macroscopic turbulence) driven by radial, inhomogeneous accretions (e.g., \citealt{Clarke2017}). 
  
    \item The radial profiles of $|\nabla v_{\mathrm{LSR}, \perp}|$ and $|\nabla v_{\mathrm{LSR}, \parallel}|$ closely resemble each other. In three of the four B5 filaments, both of these profiles are relatively flat and have median values in the range of $1.7 - 2.3$ km s$^{-1}$ pc$^{-1}$. The flat $|\nabla v_{\mathrm{LSR}, \perp}|$ profile outside of the $R_{\mathrm{flat}}$ region appears contrary to those predicted by free-fall or gravitationally dominated accretion scenarios (e.g., \citealt{Gomez2014}), which have velocity gradient profiles that increase monotonically towards the spine, without accounting for projection effects.
  
\end{enumerate}

Our $\nabla v_{\mathrm{LSR}}$ analyses of the B5 filaments revealed a wealth of complex kinematic structures at the $0.009$ pc scale, which were previously unexplored by observations and simulations. While the complex $\nabla v_{\mathrm{LSR}}$ field of B5 suggests that turbulence continues to remain important on the smaller, subsonic scales where accretion and infall motions may dominate, detailed synthetic observations of simulations are needed to determine further the nature of these motions, particularly regarding the decreasing $\sigma_{v}$ towards the filament spines. 

Even though we did not find observational trends at the smaller scales that can differentiate between filament formation models between the GT and the GHC scenarios, the flat $|\nabla v_{\mathrm{LSR}, \perp}|$ profiles we found outside of the flat inner regions of the filaments appear contrary to those naively expected from models that operate within the GHC scenario. However, realistic synthetic observations of GHC models will best determine whether or not such tension does indeed exist between these models and our observations. Further work to understand better the nature of the observed compact $|\nabla v_{\mathrm{LSR}, \perp}|$ structures, particularly in relation to the turbulent Mach number of clouds in simulations (e.g., \citealt{Clarke2017}), will also help further determining whether or not the results of this work conforms better to the GT scenarios than the GHC scenarios.
  
\acknowledgments
M.C.C. and J.D.F. acknowledge the financial support of a Discovery Grant from NSERC of Canada. The Green Bank Observatory is a facility of the National Science Foundation operated under cooperative agreement by Associated Universities, Inc. The National Radio Astronomy Observatory is a facility of the National Science Foundation operated under cooperative agreement by Associated Universities, Inc. M.C.C. further thanks Bob Ross, who taught us that there are no mistakes, only happy accidents.

\facilities{VLA(K-band), GBT(KFPA)}


\software{astropy \citep{Robitaille2013}, CRISPy \citep{ChenM2020}, LicPy \citep{Rufat2018}, MUFASA \citep{ChenM2020}, PySpeckit \citep{Ginsburg2011}}



\vskip12pt
\appendix
\section{Kinematics of the second velocity component}\label{appendix:vcomps}

\begin{figure*}
\centering
\includegraphics[width=0.95\textwidth]{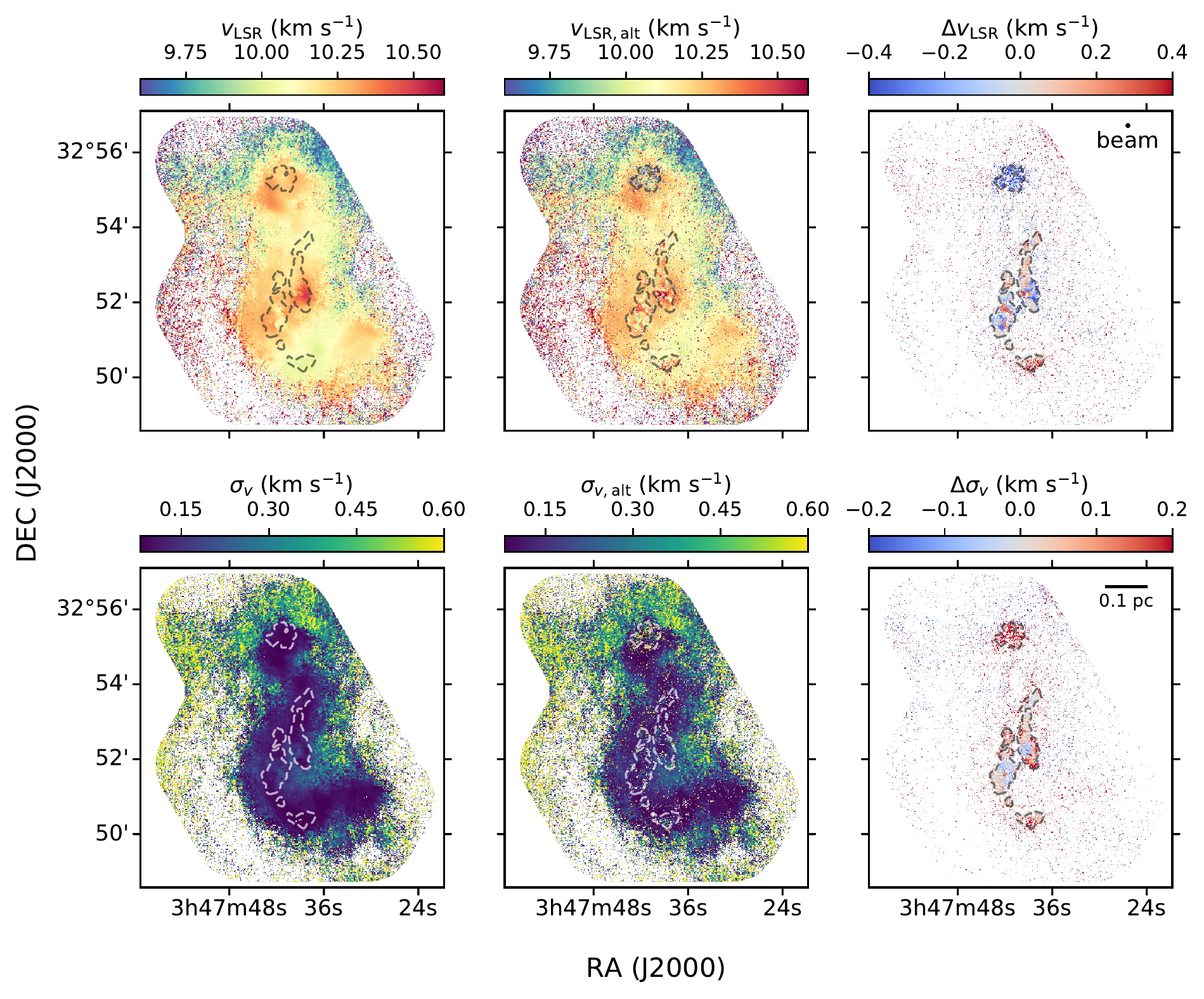}
\caption{Left column: the final maps of $v_{\mathrm{LSR}}$ (top) and $\sigma_v$ (bottom) we used for our analysis (i.e., those shown in Figure \ref{fig:spine_maps}). Center column: same as those presented in the left column but with the component we excluded from our analysis placed in pixels with $\ln{K^2_1} > 5$. Right column: the difference map produced from subtracting the left column maps from their right column counter-parts, with one-component-fitted pixels masked out. All panels are overlaid with dashed contours of where the smoothed $\ln{K^2_1}$ map is greater than 5. The beam of the data and a physical scale bar are shown in the top and bottom panels of the left column, respectively.}
\label{fig:kin_compare_two}
\end{figure*}

Figure \ref{fig:kin_compare_two} shows the final maps of $v_{\mathrm{LSR}}$ and $\sigma_v$ we used for our analysis (left column) and the alternative maps (center column) constructed with the excluded component replacing their adopted counterparts in pixels with $\ln{K^2_1} > 5$, i.e., those best-fitted by two-component models. As expected based the selection criteria described in Section \ref{subsec:vc_maps}, our final maps appear much smoother than their alternative counterparts, indicating that our final maps are indeed tracing the kinematically-coherent structure that dominates the emission observed in B5. Furthermore, the alternative map within the smoothed $\ln{K^2_1} > 5$ contours appears rather patchy, suggesting that the excluded component, unlike the adopted counterpart, does not trace another single, coherent structure. The excluded component likely consists of many compact, distinct structures. 

The right panel of Figure \ref{fig:kin_compare_two} shows the difference maps between the maps shown in the left and center panels. The $v_{\mathrm{LSR}}$ differences ($\Delta v_{\mathrm{LSR}}$) between the two components often have transonic or supersonic values (i.e., $\Delta v_{\mathrm{LSR}} > 0.2$ km s$^{-1}$), and contain compact structures that alternate in signs. Such a patchy morphology suggests the excluded component likely corresponds to clumpy, compact structures being accreted onto the filaments, such as those speculated by \citep{Clarke2017}. Future work is needed to investigate further the natures of these second component features detected in the B5 filaments.

\vskip12pt
\bibliography{bibliography}{}
\bibliographystyle{aasjournal}



\end{document}